\documentclass[11pt]{article}
\usepackage{amssymb}

\usepackage{graphicx}
\usepackage{amsmath}

\input{tcilatex}

\begin{document}

\centerline{\textbf{\Large}}

\vskip 0.8truecm

\centerline{\textbf{\Large VARIATIONAL DERIVATION OF RELATIVISTIC}} \vskip %
0.3truecm \centerline{\textbf{\Large FERMION-ANTIFERMION WAVE EQUATIONS}} %
\vskip 0.3truecm \centerline{\textbf{\Large IN QED}}

\vskip 0.6truecm \centerline{\large Andrei G. Terekidi and Jurij W. Darewych}

\vskip 0.5truecm 
\centerline{\footnotesize \emph{Department of Physics and
Astronomy, York University, Toronto, Ontario, M3J 1P3, Canada}}

\vskip 0.6truecm


\noindent \textbf{Abstract} We present a variational method for deriving
relativistic two-fermion wave equations in a Hamiltonian formulation of QED.
A reformulation of QED is performed, in which covariant Green functions are
used to solve for the electromagnetic field in terms of the fermion fields.
The resulting modified Hamiltonian contains the photon propagator directly.
The reformulation permits one to use a simple Fock-space variational trial
state to derive relativistic fermion-antifermion wave equations from the
corresponding quantum field theory. We verify that the energy eigenvalues
obtained from the wave equation agree with known results for positronium.

\vskip 0.8truecm

\noindent\textbf{\large 1. Introduction} 

\vskip0.4truecm The description of relativistic bound and quasi-bound (i.e.
unstable) few body systems continues to be an active area of research. The
traditional method of treating relativistic bound states in quantum field
theory (QFT) is by means of the Bethe-Salpeter (BS) equation. However, this
approach has a number of difficulties, including the appearance of
relative-time coordinates and negative-energy solutions. In practice, the
interaction kernels (potentials) in the BS equation are obtained from
covariant perturbation theory, which may be of questionable validity for
strongly coupled systems. In addition, the BS formalism is difficult to
implement for systems of more than two particles.

An alternative approach might be the variational method, which is
non-perturbative in principle. The variational method has not been widely
used in quantum field theory, in contrast to non-relativistic systems
describable by the Schr\"{o}dinger theory, in part because of the difficulty
of constructing realistic yet tractable trial states.

It has been pointed out in previous publications [1,2] that various models
in QFT, including QED, can be reformulated, using mediating-field Green
functions, into a form particularly convenient for variational calculations.
This approach was applied recently to the study of relativistic two-body
states in the scalar Yukawa (Wick-Cutkosky) theory [3,4,5]. In the present
paper we shall implement this approach to the realistic QED theory, where
comparison with experimentally verified results are possible. In particular,
we shall use the reformulated QED Hamiltonian to derive a relativistic
fermion-antifermion wave equation and discuss its solution.

The reformulation of the QED is presented in section 2, while the
Hamiltonian and equal time quantization are given in section 3. In section 4
we use the variational principle, with simple Fock-space trial states to
derive the relativistic fermion-antifermion equations, and present their
``partial wave'' decomposition for all possible $J^{PC}$ states. The
relativistic radial equations are presented in section 5, while their
non-relativistic and semi relativistic limits are given in 6. In section 7
the energy eigenvalues are shown to yield the correct the fine and hyperfine
structure for all states. Concluding remarks are given in section 8.

\vskip0.8truecm {\normalsize \noindent {\textbf{\large 2. Reformulation of
field equations and Lagrangian}}} {\normalsize \vskip0.4truecm }

The Lagrangian of QED is $\left( \hbar =c=1\right) $%
\begin{eqnarray}
\mathcal{L} &=&\overline{\psi }(x)\left( i\gamma ^{\mu }\partial _{\mu
}-m-e\gamma ^{\mu }A_{\mu }(x)\right) \psi (x)  \notag \\
&&-\frac{1}{4}\left( \partial _{\alpha }A_{\beta }(x)-\partial _{\beta
}A_{\alpha }(x)\right) \left( \partial ^{\alpha }A^{\beta }(x)-\partial
^{\beta }A^{\alpha }(x)\right) .
\end{eqnarray}
The corresponding Euler-Lagrange equations of motion are the coupled
Dirac-Maxwell equations, 
\begin{equation}
\left( i\gamma ^{\mu }\partial _{\mu }-m\right) \psi (x)=e\gamma ^{\mu
}A_{\mu }(x)\psi (x),
\end{equation}
and 
\begin{equation}
\partial _{\mu }\partial ^{\mu }A^{\nu }(x)-\partial ^{\nu }\partial _{\mu
}A^{\mu }(x)=j^{\nu }(x),
\end{equation}
where 
\begin{equation}
j^{\nu }(x)=e\overline{\psi }(x)\gamma ^{\nu }\psi (x).
\end{equation}
The equations (2)-(3) can be decoupled in part by using the well-known
formal solution [6,7] of the Maxwell equation (3), namely 
\begin{equation}
A_{\mu }(x)=A_{\mu }^{0}(x)+\int d^{4}x^{\prime }D_{\mu \nu }(x-x^{\prime
})j^{\nu }(x^{\prime }),
\end{equation}
where $D_{\mu \nu }(x-x^{\prime })$ is a Green function (or photon
propagator in QFT terminology), defined by 
\begin{equation}
\partial _{\alpha }\partial ^{\alpha }D_{\mu \nu }(x-x^{\prime })-\partial
_{\mu }\partial ^{\alpha }D_{\alpha \nu }(x-x^{\prime })=g_{\mu \nu }\delta
^{4}(x-x^{\prime }),
\end{equation}
and $A_{\mu }^{0}(x)$ is a solution of the homogeneous (or ``free field'')
equation (3) with $j^{\mu }(x)=0.$

We recall, in passing, that equation (6) does not define the covariant Green
function $D_{\mu \nu }(x-x^{\prime })$ uniquely. For one thing, one can
always add a solution of the homogeneous equation (eq. (6) with $g_{\mu \nu
}\rightarrow 0$). This allows for a certain freedom in the choice of $D_{\mu
\nu }$, as is discussed in standard texts (e.g. ref. [6,7]). In practice,
the solution of eq. (6), like that of eq. (3), requires a choice of gauge.
However, we do not need to specify one at this stage.

Substitution of the formal solution (5) into equation (2) yields the
``partly reduced'' equations,

\begin{equation}
\left( i\gamma ^{\mu }\partial _{\mu }-m\right) \psi (x)=e\gamma ^{\mu
}\left( A_{\mu }^{0}(x)+\int d^{4}x^{\prime }D_{\mu \nu }(x-x^{\prime
})j^{\nu }(x^{\prime })\right) \psi (x),
\end{equation}
which is a nonlinear Dirac equation. To our knowledge no exact (analytic or
numeric) solution of equation (7) for classical fields have been reported in
the literature. However, approximate solutions have been discussed by
various authors, particularly Barut and his co-workers (see ref. [8,9] and
citations therein). In any case, our interest here is in the quantized field
theory.

The partially reduced equation (7) is derivable from the stationary action
principle 
\begin{equation}
\delta S\left[ \psi \right] =\delta \int d^{4}x\mathcal{L}_{R}=0
\end{equation}
with the Lagrangian density 
\begin{equation}
\mathcal{L}_{R}=\overline{\psi }(x)\left( i\gamma ^{\mu }\partial _{\mu
}-m-e\gamma _{\mu }A_{0}^{\mu }(x)\right) \psi (x)-\frac{1}{2}\int
d^{4}x^{\prime }j^{\mu }(x^{\prime })D_{\mu \nu }(x-x^{\prime })j^{\nu }(x)
\end{equation}
provided that the Green function is symmetric in the sense that 
\begin{equation}
D_{\mu \nu }(x-x^{\prime })=D_{\mu \nu }(x^{\prime }-x)\;\;\;\;\;and\;\;\
\;\ D_{\mu \nu }(x-x^{\prime })=D_{\nu \mu }(x-x^{\prime }).
\end{equation}

One can proceed to do conventional covariant perturbation theory using the
reformulated QED Lagrangian (9). The interaction part of (9) has a somewhat
modified structure from that of the usual formulation of QED. Thus, there
are two interaction terms. The last term of (9) is a ``current-current''
interaction which contains the photon propagator sandwiched between the
fermionic currents. As such, it corresponds to Feynman diagrams without
external photon lines. The term containing $A_{0}^{\mu }$ corresponds to
diagrams that cannot be generated by the term containing $D_{\mu \nu }$,
including diagrams involving external photon lines (care would have to be
taken not to double count physical effects). However, we shall not pursue
covariant perturbation theory in this work. Rather, we shall consider a
variational approach that allows one to derive relativistic few-fermion
equations, and to study their bound and scattering solutions.


{\normalsize \vskip0.8truecm \noindent {\textbf{\large 3. Hamiltonian of the
quantized theory in the equal-time formalism}} }

{\normalsize \vskip 0.4truecm }

{\normalsize 
}We consider this theory in the quantized, equal-time formalism. To this end
we write down the Hamiltonian density corresponding to the Lagrangian (9),
with the term for the free $A_{0}^{\mu }(x)$ field suppressed since it will
not contribute to the results presented in this paper. The relevant
expression is: 
\begin{equation}
\mathcal{H}_{R}=\mathcal{H}_{0}\mathcal{+H}_{I},
\end{equation}
where

\begin{eqnarray}
\mathcal{H}_{0} &=&\psi ^{\dagger }(x)\left( -i\overrightarrow{\alpha }\cdot
\nabla +m\beta \right) \psi (x), \\
\mathcal{H}_{I} &=&\frac{1}{2}\int d^{4}x^{\prime }j^{\mu }(x^{\prime
})D_{\mu \nu }(x-x^{\prime })j^{\nu }(x),
\end{eqnarray}

We construct a quantized theory by the imposition of anticommutation rules
for the fermion fields, namely 
\begin{equation}
\left\{ \psi _{\alpha }(\mathbf{x},t),\psi _{\beta }^{\dagger }(\mathbf{y}%
,t)\right\} =\delta _{\alpha \beta }\delta ^{3}\left( \mathbf{x}-\mathbf{y}%
\right) ,
\end{equation}
while all other vanish. In addition, if $A_{0}^{\mu }\neq 0,$ there would be
the usual commutation rules for the $A_{0}^{\mu }$ field, and commutation of
the $A_{0}^{\mu }$ field operators with the $\psi $ field operators.

To specify our notation, we quote the usual Fourier decomposition of the
field operators, namely 
\begin{equation}
\psi (x)=\sum_{s}\int \frac{d^{3}p}{\left( 2\pi \right) ^{3/2}}\left( \frac{m%
}{\omega _{p}}\right) ^{1/2}\left[ b_{\mathbf{p}s}u\left( \mathbf{p}%
,s\right) e^{-ip\cdot x}+d_{\mathbf{p}s}^{\dagger }v\left( \mathbf{p}%
,s\right) e^{ip\cdot x}\right] ,
\end{equation}
with $p=p^{\mu }=\left( \omega _{p},\mathbf{p}\right) $, and $\omega _{p}=%
\sqrt{m^{2}+\mathbf{p}^{2}}$. Dirac spinors $u$\ and $v$ for free particles
of mass $m$,\ where $\left( \gamma ^{\mu }{p}_{\mu }-m\right) u\left( 
\mathbf{p},s\right) =0$,\ $\left( \gamma ^{\mu }{p}_{\mu }+m\right) v\left( 
\mathbf{p},s\right) =0$, are normalized such that 
\begin{eqnarray}
u^{\dagger }\left( \mathbf{p},s\right) u\left( \mathbf{p},\sigma \right)
&=&v^{\dagger }\left( \mathbf{p},s\right) v\left( \mathbf{p},\sigma \right) =%
\frac{\omega _{p}}{m}\delta _{s\sigma }, \\
u^{\dagger }\left( \mathbf{p},s\right) v\left( \mathbf{p},\sigma \right)
&=&v^{\dagger }\left( \mathbf{p},s\right) u\left( \mathbf{p},\sigma \right)
=0.
\end{eqnarray}

The creation and annihilation operators $b^{\dagger }$, $b$ of the (free)
fermions of mass $m$, and $d^{\dagger }$, $d$ for the corresponding
antifermions, satisfy the usual anticommutation relations. The non-vanishing
ones are 
\begin{equation}
\left\{ b_{\mathbf{p}s},b_{\mathbf{q}\sigma }^{\dagger }\right\} =\left\{ d_{%
\mathbf{p}s},d_{\mathbf{q}\sigma }^{\dagger }\right\} =\delta _{s\sigma
}\delta ^{3}\left( \mathbf{p}-\mathbf{q}\right) .
\end{equation}


{\normalsize \vskip 0.8truecm \noindent{\textbf{\large 4. Variational
principle and fermion-antifermion trial states}} }

{\normalsize \vskip 0.4truecm }

{\normalsize 
}Unfortunately we do not know how to obtain exact eigenstates of the
Hamiltonian (11). Therefore we shall resort to a variational approximation,
based on the variational principle 
\begin{equation}
\delta \left\langle \psi \right| \widehat{H}-E\left| \psi \right\rangle
_{t=0}=0.
\end{equation}
For a fermion-antifermion system , the simplest Fock-space trial state that
can be written down in the rest frame is

\begin{equation}
\left| \psi _{T}\right\rangle =\underset{s_{1}s_{2}}{\sum }\int d^{3}\mathbf{%
p}F_{s_{1}s_{2}}(\mathbf{p})b_{\mathbf{p}s_{1}}^{\dagger }d_{-\mathbf{p}%
s_{2}}^{\dagger }\left| 0\right\rangle ,
\end{equation}
where $F_{s_{1}s_{2}}$ are four adjustable functions. We use this trial
state to evaluate the matrix elements needed to implement the variational
principle (19), namely 
\begin{equation}
\left\langle \psi _{T}\right| :\widehat{H}_{0}-E:\left| \psi
_{T}\right\rangle =\underset{s_{1}s_{2}}{\sum }\int d^{3}\mathbf{p}%
F_{s_{1}s_{2}}^{\ast }(\mathbf{p})F_{s_{1}s_{2}}(\mathbf{p})\left( 2\omega
_{p}-E\right)
\end{equation}
and

\begin{equation*}
\left\langle \psi _{T}\right| :\widehat{H}_{I}:\left| \psi _{T}\right\rangle
=\frac{e^{2}m^{2}}{\left( 2\pi \right) ^{3}}\underset{s_{1}^{\prime
}s_{2}^{\prime }s_{1}s_{2}}{\sum }\int \frac{d^{3}\mathbf{p}d^{3}\mathbf{p}%
^{\prime }}{\omega _{p}\omega _{p^{\prime }}}F_{s_{1}s_{2}}^{\ast }(\mathbf{p%
})F_{s_{1}^{\prime }s_{2}^{\prime }}(\mathbf{p}^{\prime })\times
\end{equation*}

\begin{equation}
\times \left( 
\begin{array}{c}
-\overline{u}\left( \mathbf{p},s_{1}\right) \gamma ^{\mu }u\left( \mathbf{p}%
^{\prime },s_{1}^{\prime }\right) D_{\mu \nu }(p-p^{\prime })\overline{v}%
\left( -\mathbf{p}^{\prime },s_{2}^{\prime }\right) \gamma ^{\nu }v\left( -%
\mathbf{p,}s_{2}\right) \\ 
+\overline{u}\left( \mathbf{p},s_{1}\right) \gamma ^{\mu }v\left( -\mathbf{p,%
}s_{2}\right) D_{\mu \nu }\left( p+p^{\prime }\right) \overline{v}\left( -%
\mathbf{p}^{\prime },s_{2}^{\prime }\right) \gamma ^{\nu }u\left( \mathbf{p}%
^{\prime },s_{1}^{\prime }\right)
\end{array}
\right) \,,
\end{equation}
where $p=(\omega _{p},{}\mathbf{p})$, $p^{\prime }=(\omega _{p^{\prime }},{}%
\mathbf{p}^{\prime })$, with ${}\mathbf{p}+\mathbf{p}^{\prime }=0$ (i.e. $%
p+p^{\prime }=(2\omega _{p},0)$) in the rest frame, and 
\begin{equation}
D_{\mu \nu }(x-x^{\prime })=\int \frac{d^{4}k}{(2\pi )^{4}}D_{\mu \nu
}(k)e^{-ik\cdot (x-x^{\prime })}.
\end{equation}

We have normal-order the entire Hamiltonian, since this circumvents the need
for mass renormalization which would otherwise arise. Not that there is
difficulty with handling mass renormalization in the present formalism (as
shown in various earlier papers; see, for example, ref. [10] and citations
therein). It is simply that we are not interested in mass renormalization
here, since it has no effect on the two-body bound state energies that we
obtain in this paper. Furthermore, the approximate trial state (20), which
we use in this work, is incapable of sampling loop effects. Thus, the
normal-ordering of the entire Hamiltonian does not ``sweep under the
carpet'' loop effects, since none arise at the present level of
approximation, that is with the trial state $\mid \psi _{T}\rangle $
specified in eq. (20) .

The variational principle (19) leads to the following equation 
\begin{eqnarray}
&&\sum_{s_{1}s_{2}}\int d^{3}\mathbf{p}\left( 2\omega _{p}-E\right)
F_{s_{1}s_{2}}(\mathbf{p})\delta F_{s_{1}s_{2}}^{\ast }(\mathbf{p})  \notag
\\
&&-\frac{m^{2}}{\left( 2\pi \right) ^{3}}\underset{\sigma _{1}\sigma
_{2}s_{1}s_{2}}{\sum }\int \frac{d^{3}\mathbf{p}d^{3}\mathbf{q}}{\omega
_{p}\omega _{q}}F_{\sigma _{1}\sigma _{2}}(\mathbf{q})\left( -i\right) 
\mathcal{M}_{s_{1}s_{2}\sigma _{1}\sigma _{2}}\left( \mathbf{p,q}\right)
\delta F_{s_{1}s_{2}}^{\ast }(\mathbf{p})=0,
\end{eqnarray}
where $\mathcal{M}_{s_{1}s_{2}\sigma _{1}\sigma _{2}}\left( \mathbf{p,q}%
\right) $ is an invariant ``matrix element'', which contains two terms: 
\begin{equation}
\mathcal{M}_{s_{1}s_{2}\sigma _{1}\sigma _{2}}\left( \mathbf{p,q}\right) =%
\mathcal{M}_{s_{1}s_{2}\sigma _{1}\sigma _{2}}^{ope}\left( \mathbf{p,q}%
\right) +\mathcal{M}_{s_{1}s_{2}\sigma _{1}\sigma _{2}}^{ann}\left( \mathbf{%
p,q}\right) ,
\end{equation}
where 
\begin{equation}
\mathcal{M}_{s_{1}s_{2}\sigma _{1}\sigma _{2}}^{ope}\left( \mathbf{p,q}%
\right) =-\overline{u}\left( \mathbf{p},s_{1}\right) \left( -ie\gamma ^{\mu
}\right) u\left( \mathbf{q},\sigma _{1}\right) iD_{\mu \nu }(p-q)\overline{v}%
\left( -\mathbf{q},\sigma _{2}\right) \left( -ie\gamma ^{\nu }\right)
v\left( -\mathbf{p},s_{2}\right) ,
\end{equation}
\begin{equation}
\mathcal{M}_{s_{1}s_{2}\sigma _{1}\sigma _{2}}^{ann}\left( \mathbf{p,q}%
\right) =\overline{u}\left( \mathbf{p},s_{1}\right) \left( -ie\gamma ^{\mu
}\right) v\left( -\mathbf{p},s_{2}\right) iD_{\mu \nu }\left( p+q\right) 
\overline{v}\left( -\mathbf{q},\sigma _{2}\right) \left( -ie\gamma ^{\nu
}\right) u\left( \mathbf{q},\sigma _{1}\right) \,,
\end{equation}
correspond to the usual one-photon exchange and virtual annihilation Feynman
diagrams.

At this point it is worthwhile to make a few comments about our equation
(24) and to compare its general features with other two-fermion equations,
particularly field-theory based approaches. Firstly we note that the present
variational derivation leads to momentum-space Salpeter-like equations, with
at most four independent components $F_{s_{1}s_{2}}\left( \mathbf{p}\right) $%
. The equations have only positive-energy solutions, as is evident from eq.
(24) with the interaction turned off, in which case only $E=2\omega _{p}>0$
is obtained. This is in contrast to the BS equation, which is a 16-component
equation and contains both positive, negative and mixed energy solutions.

The interaction kernels, represented by the covariant $\mathcal{M}$%
-matrices, result from the variational derivation, that is, they are not put
in by hands. This is in contrast to two fermion equations, which are not
derived from a underlying quantum field theory, such as various two-body
generalizations of the one-body Dirac equation. There are many such
equations on the market, for example the eight component two-fermion
equation of Pilkuhn [11]. In these treatments QFT effects, such as the
virtual annihilation interaction (eq. (27)) do not arise naturally but need
to be added in.

The fact that only the lowest order (``tree level'') diagrams appear in our
eq. (24) is a reflection of the fact that we have used the simplest possible
variational ansatz (20). Even so, it is important to note that, because of
the reformulation discussed in section 2 and 3, their derivation does not
require additional Fock-space terms in the variational state (20) as is the
case in traditional (non-reformulated) treatments (e.g. [12]-[14]).

In the non-relativistic limit, the functions $F_{s_{1}s_{2}}$ can be written
as 
\begin{equation}
F_{s_{1}s_{2}}(\mathbf{p})=F(\mathbf{p})\Lambda _{s_{1}s_{2}},
\end{equation}
where the non-zero elements of $\Lambda _{ij}$ for total spin singlet ($S=0$%
) states are $\Lambda _{12}=-\Lambda _{21}=\frac{1}{\sqrt{2}}$, while for
the spin triplet ($S=1$)\ states the non-zero elements are $\Lambda _{11}=1$
for $m_{s}=+1,$ $\Lambda _{12}=\Lambda _{21}=\frac{1}{\sqrt{2}}$ for $%
m_{s}=0 $, and $\Lambda _{22}=1$ for $m_{s}=-1$. We use the notation that
the subscripts $1$ and 2 of $\Lambda $ correspond to $m_{s}=1/2$ and $%
m_{s}=-1/2$ (or $\uparrow $\ and\ $\downarrow $) respectively. Substituting
(28) into (24), the variational procedure, after multiplying the result by $%
\Lambda _{s_{1}s_{2}}$ and summing over $s_{1}$ and $s_{2}$, gives the
equation 
\begin{equation}
(2\omega _{p}-E)F(\mathbf{p})=\frac{1}{(2\pi )^{3}}\int d^{3}\mathbf{q}\,%
\mathcal{K}(\mathbf{p},\mathbf{q})F(\mathbf{q}),
\end{equation}
where 
\begin{equation}
\mathcal{K}(\mathbf{p},\mathbf{q})=-i\frac{m^{2}}{\omega _{p}\omega _{q}}%
\sum_{s_{1}s_{2}\sigma _{1}\sigma _{2}}\Lambda _{s_{1}s_{2}}\mathcal{M}%
_{s_{1}s_{2}\sigma _{1}\sigma _{2}}\left( \mathbf{p,q}\right) \Lambda
_{\sigma _{1}\sigma _{2}}.
\end{equation}
To lowest-order in $|\mathbf{p}|/m$ (i.e. in the non-relativistic limit),
the kernel (30) reduces to $\mathcal{K}=e^{2}/\left| \mathbf{p-q}\right|
^{2} $, and so (29) reduces to the (momentum-space) Schr\"{o}dinger equation 
\begin{equation}
\left( \frac{\mathbf{p}^{2}}{2\mu }-\varepsilon \right) F(\mathbf{p})=\frac{1%
}{(2\pi )^{3}}\int d^{3}\,\mathbf{q}\frac{e^{2}}{\left| \mathbf{p-q}\right|
^{2}}F(\mathbf{q}),
\end{equation}
where $\varepsilon =E-2m$ and $\mu =m/2$. This verifies that the
relativistic two-fermion equation (24) has the expected non-relativistic
limit.

In the relativistic case we do not complete the variational procedure in
(24) at this stage to get equations for the four adjustable functions $%
F_{s_{1}s_{2}}$, because they are not independent in general. Indeed we
require that the trial state be an eigenstate of the total angular momentum
operator (in relativistic form), its projection, parity and charge
conjugation, namely that 
\begin{equation}
\left[ 
\begin{array}{c}
\widehat{\mathbf{J}}^{2} \\ 
\widehat{J}_{3} \\ 
\widehat{\mathcal{P}} \\ 
\widehat{\mathcal{C}}
\end{array}
\right] \left| \psi _{T}\right\rangle =\left[ 
\begin{array}{c}
J\left( J+1\right) \\ 
m_{J} \\ 
P \\ 
C
\end{array}
\right] \left| \psi _{T}\right\rangle ,
\end{equation}
where $m_{J}=J,J-1,...,-J$ as usual. We present explicit forms for the
operators $\widehat{\mathbf{J}}^{2}$, $\widehat{J}_{3}$ in Appendix A. The
form for $\widehat{\mathbf{J}}^{2}$, eq. (109), in particular, is not
readily found in standard texts and reference books.

The functions $F_{s_{1}s_{2}}(\mathbf{p})$ can be written in the general
form 
\begin{equation}
F_{s_{1}s_{2}}(\mathbf{p})=\sum_{\ell
_{s_{1}s_{2}}}\sum_{m_{s_{1}s_{2}}}f_{s_{1}s_{2}}^{\ell
_{s_{1}s_{2}}m_{s_{1}s_{2}}}\left( p\right) Y_{\ell
_{s_{1}s_{2}}}^{m_{s_{1}s_{2}}}(\hat{\mathbf{p}}),
\end{equation}
where $Y_{\ell _{s_{1}s_{2}}}^{m_{s_{1}s_{2}}}(\hat{\mathbf{p}})$\ are the
usual spherical harmonics. Here and henceforth we will use the notation $%
p=\left| \mathbf{p}\right| $ etc. (four-vectors will be written as $p^{\mu }$%
). The orbital indexes $\ell _{s_{1}s_{2}}$and $m_{s_{1}s_{2}}$ depend on
the spin indexes $s_{1}$ and $s_{2}$\ and are specified by equations (32).
The radial coefficients $f_{s_{1}s_{2}}^{\ell
_{s_{1}s_{2}}m_{s_{1}s_{2}}}\left( p\right) $ in the expansion (33) also
depend on the spin variables.

Substitution of (33) into (20) and then into (32) leads to two categories of
relations among the adjustable functions, as shown in Appendices A and B. It
follows that, for trial states of the form (20), the total spin of the
system is a good quantum number, and the states of the system separate into
singlet states with the total spin $S=0$ (parastates) and into triplet
states with $S=1$ (orthostates). We should point out that this phenomenon is
characteristic of the fermion antifermion systems, which are charge
conjugation eigenstates, and does not arise for systems like $\mu ^{+}e^{-}$.

\vskip .2cm \noindent \textbf{The singlet states}

In this case $\ell _{s_{1}s_{2}}\equiv \ell =J,\;m_{11}=m_{22}=0\ $and$%
\;m_{12}=m_{21}=m_{J}$.\ The nonzero components of $F_{s_{1}s_{2}}(\mathbf{p}%
)$ are $F_{\uparrow \downarrow }(\mathbf{p})\equiv F_{12}(\mathbf{p}%
),\;F_{\downarrow \uparrow }(\mathbf{p})\equiv F_{21}(\mathbf{p})$\ and have
the form 
\begin{equation}
F_{s_{1}s_{2}}(\mathbf{p})=f_{s_{1}s_{2}}^{\left( sgl\right)
J}(p)Y_{J}^{m_{s_{1}s_{2}}}(\hat{\mathbf{p}}),
\end{equation}
where the relations between \ $f_{12}^{\left( sgl\right) J}(p)\;$and$%
\;\,f_{21}^{\left( sgl\right) J}(p)$ involve the Clebsch-Gordan (C-G)
coefficients $C_{Jm_{J}}^{\left( sgl\right) Jm_{s_{1}s_{2}}}$, that is 
\begin{equation}
f_{s_{1}s_{2}}^{\left( sgl\right) J}(p)=C_{Jm_{J}}^{\left( sgl\right)
Jm_{s_{1}s_{2}}}f^{J}(p),
\end{equation}
as it shown in Appendix A. We see that the spin and radial variables
separate for the singlet states in the sense that the factors $%
f_{s_{1}s_{2}}^{\left( sgl\right) J}(p)$ have a common radial function $%
f^{J}(p)$. Thus, for the singlet states we obtain 
\begin{equation}
F_{s_{1}s_{2}}(\mathbf{p})=C_{Jm_{J}}^{\left( sgl\right)
Jm_{s_{1}s_{2}}}f^{J}(p)Y_{J}^{m_{J}}(\hat{\mathbf{p}}).
\end{equation}
The C-G coefficients $C_{Jm_{J}}^{\left( sgl\right) Jm_{s_{1}s_{2}}}$\ have
a simple form: $C_{Jm_{J}}^{\left( sgl\right) Jm_{11}}=C_{Jm_{J}}^{\left(
sgl\right) Jm_{22}}=0$,\ $C_{Jm_{J}}^{\left( sgl\right)
Jm_{12}}=-C_{Jm_{J}}^{\left( sgl\right) Jm_{21}}=1$ (see Appendix A).\
Therefore for the singlet states we can write expression (20) in the
explicit form 
\begin{equation}
\left| \psi _{T}\right\rangle =\int d^{3}\mathbf{p}f^{J}(p)Y_{J}^{m_{J}}(%
\hat{\mathbf{p}})\left( b_{\mathbf{p}\uparrow }^{\dagger }d_{-\mathbf{p}%
\downarrow }^{\dagger }-b_{\mathbf{p}\downarrow }^{\dagger }d_{-\mathbf{p}%
\uparrow }^{\dagger }\right) \left| 0\right\rangle .
\end{equation}
These states are characterized by the quantum numbers $J,m_{J}$ parity $%
P=(-1)^{J+1}$ and charge conjugation $C=\left( -1\right) ^{J}$. As we can
see, the quantum numbers $\ell $ (orbital angular momentum), and total spin $%
S$ are good quantum numbers for the singlet states as well. The
spectroscopical notation is $^{1}J_{J}$.

\vskip .2cm \noindent \textbf{The triplet states}

The solution of the system (32) for $S=1$ leads to two cases (Appendix A),
namely $\ell _{s_{1}s_{2}}\equiv \ell =J$, for which

\begin{equation}
F_{s_{1}s_{2}}(\mathbf{p})=f_{s_{1}s_{2}}^{\left( tr\right)
J}(p)Y_{J}^{m_{s_{1}s_{2}}}(\hat{\mathbf{p}}),
\end{equation}
and $\ell _{s_{1}s_{2}}\equiv \ell =J\mp 1$, for which

\begin{equation}
F_{s_{1}s_{2}}(\mathbf{p})=f_{s_{1}s_{2}}^{J-1}(p)Y_{J-1}^{m_{s_{1}s_{2}}}(%
\hat{\mathbf{p}})+f_{s_{1}s_{2}}^{J+1}(p)Y_{J+1}^{m_{s_{1}s_{2}}}(\hat{%
\mathbf{p}}),
\end{equation}
where 
\begin{equation}
m_{11}=m_{J}-1,\;\ \ \ m_{12}=m_{21}=m_{J},\;\ \ \ m_{22}=m_{J}+1.
\end{equation}
The expressions for $f_{s_{1}s_{2}}^{\ell }(p)$ in both cases involve the
C-G coefficients $C_{Jm_{J}}^{\left( tr\right) \ell m_{s}}$ for $S=1$ listed
in Appendix A, that is 
\begin{equation}
f_{s_{1}s_{2}}^{\left( tr\right) \ell }(p)=C_{Jm_{J}}^{\left( tr\right) \ell
m_{s}}f^{\ell }(p),
\end{equation}
where the index $m_{s}$ is defined as 
\begin{eqnarray}
m_{s} &=&+1,\;\ \ when\;\ \ m_{s_{1}s_{2}}=m_{11},  \notag \\
m_{s} &=&0,\;\ \ \ \ when\;\ \ \ m_{s_{1}s_{2}}=m_{12}=m_{21}, \\
m_{s} &=&-1,\;\ \ when\;\ \ m_{s_{1}s_{2}}=m_{22}.  \notag
\end{eqnarray}

Thus, for the triplet states with $\ell =J$%
\begin{equation}
F_{s_{1}s_{2}}(\mathbf{p})=C_{Jm_{J}}^{\left( tr\right)
Jm_{s}}f^{J}(p)Y_{J}^{m_{s_{1}s_{2}}}(\hat{\mathbf{p}}).
\end{equation}
These functions correspond to states, which can be characterized by the
quantum numbers $J,m_{J}$, parity $P=(-1)^{J+1}$ and charge conjugation $%
C=\left( -1\right) ^{J+1}$. The orbital angular momentum $\ell $, as well as
the total spin $S=1$, are good quantum numbers in this case. The
spectroscopic notation for these states is $^{3}J_{J}$.

For the triplet states with $\ell =J\mp 1$ we obtain the result 
\begin{equation}
F_{s_{1}s_{2}}(\mathbf{p})=C_{Jm_{J}}^{\left( tr\right)
(J-1)m_{s}}f^{J-1}(p)Y_{J-1}^{m_{s_{1}s_{2}}}(\hat{\mathbf{p}}%
)+C_{Jm_{J}}^{\left( tr\right) (J+1)m_{s}}f^{J+1}(p)Y_{J+1}^{m_{s_{1}s_{2}}}(%
\hat{\mathbf{p}}),
\end{equation}
which involves two radial functions $f^{J-1}(p)$ and $f^{J+1}(p)$
corresponding to $\ell =J-1$ and $\ell =J+1$. This means that $\ell $ is not
a good quantum number. Such states are characterized by quantum numbers $J,$ 
$m_{J},$ $P=(-1)^{J}$, charge conjugation $C=\left( -1\right) ^{J}$ and spin 
$S=1$. In spectroscopic notation, these states are a mixture of $^{3}\left(
J-1\right) _{J}$ and $^{3}\left( J+1\right) _{J}$\ states.

The requirement that the states be charge conjugation eigenstates (the last
equation of (32)) is intimately tied to the conservation of total spin.
Indeed, a linear combination of singlet and triplet states like 
\begin{equation}
F_{s_{1}s_{2}}(\mathbf{p})=C_{1}f_{s_{1}s_{2}}^{\left( sgl\right)
J}(p)Y_{J}^{m_{s_{1}s_{2}}}(\hat{\mathbf{p}})+C_{2}f_{s_{1}s_{2}}^{\left(
tr\right) J}(p)Y_{J}^{m_{s_{1}s_{2}}}(\hat{\mathbf{p}}),
\end{equation}
satisfies the first three equations of (32). However, it is unacceptable for
describing a fermion-antifermion system because the first and the second
terms in (45) have different charge conjugation. For a system of two
particles of different mass (such as $\mu ^{+}e^{-}$) charge conjugation is
not applicable, so that the total spin would not be conserved.


{\normalsize \vskip0.8truecm \noindent {\textbf{\large 5. The relativistic
radial equations and application to positronium-like systems}} }

{\normalsize \vskip 0.4truecm }

{\normalsize 
}We return to equation (24) and replace the functions $F_{s_{1}s_{2}}(p)$ by
the expression (36) for singlet states and by (43) and (44) for triplet
states. The variational procedure then leads to the following results:

For the singlet states $\ell =J$, $P=(-1)^{J+1}$, $C=(-1)^{J}$, the radial
equations are

{\normalsize 
\begin{equation}
\left( 2\omega _{p}-E\right) f^{J}(p)=\frac{m^{2}}{\left( 2\pi \right) ^{3}}%
\int \frac{q^{2}dq}{\omega _{p}\omega _{q}}\mathcal{K}^{\left( sgl\right)
}\left( p,q\right) f^{J}(q),
\end{equation}
}where the kernel 
\begin{equation}
\mathcal{K}^{\left( sgl\right) }\left( p,q\right) =\underset{%
s_{1}s_{2}\sigma _{1}\sigma _{2}m_{J}}{-i\sum }\int d\hat{\mathbf{p}}\,d\hat{%
\mathbf{q}}\,C_{Jm_{J}}^{\left( sgl\right) s_{1}s_{2}\sigma _{1}\sigma _{2}}%
\mathcal{M}_{s_{1}s_{2}\sigma _{1}\sigma _{2}}\left( \mathbf{p,q}\right)
Y_{J}^{m_{J}\ast }(\hat{\mathbf{p}})Y_{J}^{m_{J}}(\hat{\mathbf{q}}),
\end{equation}
is defined by the invariant $M$-matrix and the coefficients 
\begin{equation}
C_{Jm_{J}}^{\left( sgl\right) s_{1}s_{2}\sigma _{1}\sigma _{2}}\equiv
C_{Jm_{J}}^{\left( sgl\right) Jm_{\sigma }}C_{Jm_{J}}^{\left( sgl\right)
Jm_{s}}/\sum_{\nu _{1}\nu _{2}m_{J}}\left( C_{Jm_{J}}^{\left( sgl\right)
Jm_{\nu }}\right) ^{2}.
\end{equation}
Here we have summed over $m_{J}$, because of the $(2J+1)$-fold energy
degeneracy.

For the triplet states, we obtain different equations for the $\ell =J$, and 
$\ell =J\mp 1$ cases. Thus for the states with $\ell =J$, $P=(-1)^{J+1}$, $%
C=(-1)^{J+1}$ the result is 
\begin{equation}
\left( 2\omega _{p}-E\right) f^{J}(p)=\frac{m^{2}}{\left( 2\pi \right) ^{3}}%
\int \frac{q^{2}dq}{\omega _{p}\omega _{q}}\mathcal{K}^{\left( tr\right)
}(p,q)f^{J}(q),
\end{equation}
where the kernel $K^{\left( tr\right) }$\ is formally like that of (47),
namely,

\begin{equation}
\mathcal{K}^{\left( tr\right) }(p,q)=\underset{s_{1}s_{2}\sigma _{1}\sigma
_{2}m_{J}}{-i\sum }C_{Jm_{J}}^{\left( tr\right) s_{1}s_{2}\sigma _{1}\sigma
_{2}}\int d\hat{\mathbf{p}}\,d\hat{\mathbf{q}}\,\mathcal{M}%
_{s_{1}s_{2}\sigma _{1}\sigma _{2}}\left( \mathbf{p,q}\right)
Y_{J}^{m_{s_{1}s_{2}}\ast }(\hat{\mathbf{p}})Y_{J}^{m_{\sigma _{1}\sigma
_{2}}}(\hat{\mathbf{q}}).
\end{equation}
However it involves different C-G coefficients, namely

{\normalsize 
\begin{equation}
C_{Jm_{J}}^{\left( tr\right) s_{1}s_{2}\sigma _{1}\sigma
_{2}}=C_{Jm_{J}}^{\left( tr\right) Jm_{\sigma }}C_{Jm_{J}}^{\left( tr\right)
Jm_{s}}/\sum_{\nu _{1}\nu _{2}m_{J}}\left( C_{Jm_{J}}^{\left( tr\right)
Jm_{\nu }}\right) ^{2}.
\end{equation}
}

For the triplet states with $\ell =J\mp 1$, we have two independent radial
functions $f^{J-1}(p)\;$and$\;f^{J+1}(p)$.\ Thus the variational equation
(24) leads to a system of coupled equations for $f^{J-1}(p)\;$and$%
\;f^{J+1}(p)$. It is convenient to write them in matrix form,

\begin{equation}
\left( 2\omega _{p}-E\right) \mathbb{F}\left( p\right) =\frac{m^{2}}{\left(
2\pi \right) ^{3}}\int \frac{q^{2}dq}{\omega _{p}\omega _{q}}\mathbb{K}%
\left( p,q\right) \mathbb{F}\left( q\right) ,
\end{equation}
where 
\begin{equation}
\mathbb{F}\left( p\right) =\left[ 
\begin{array}{c}
f^{J-1}(p) \\ 
f^{J+1}(p)
\end{array}
\right] ,
\end{equation}
and 
\begin{equation}
\mathbb{K}\left( p,q\right) =\left[ 
\begin{array}{cc}
\mathcal{K}_{11}\left( p,q\right) & \mathcal{K}_{12}\left( p,q\right) \\ 
\mathcal{K}_{21}\left( p,q\right) & \mathcal{K}_{22}\left( p,q\right)
\end{array}
\right] .
\end{equation}
The kernels $K_{ij}$ are similar in form to (47) and (50), that is 
\begin{equation}
\mathcal{K}_{ij}\left( p,q\right) =\underset{\sigma _{1}\sigma
_{2}s_{1}s_{2}m_{J}}{-i\sum }C_{Jm_{J}ij}^{s_{1}s_{2}\sigma _{1}\sigma
_{2}}\int d\hat{\mathbf{p}}\,d\hat{\mathbf{q}}\,\mathcal{M}%
_{s_{1}s_{2}\sigma _{1}\sigma _{2}}\left( \mathbf{p,q}\right) Y_{\ell
_{j}}^{m_{\sigma _{1}\sigma _{2}}}(\hat{\mathbf{q}})Y_{\ell
_{i}}^{m_{s_{1}s_{2}}\ast }(\hat{\mathbf{p}}).
\end{equation}
However the coefficients $C_{Jm_{J}ij}^{s_{1}s_{2}\sigma _{1}\sigma _{2}}$
are defined by expression

\begin{equation}
C_{Jm_{J}ij}^{s_{1}s_{2}\sigma _{1}\sigma _{2}}=C_{Jm_{J}}^{\left( tr\right)
\ell _{j}m_{\sigma }}C_{Jm_{J}}^{\left( tr\right) \ell
_{i}m_{s}}/\sum_{s_{1}s_{2}m_{J}}\left( C_{Jm_{J}}^{\left( tr\right) \ell
_{i}m_{s}}\right) ^{2},
\end{equation}
where $\ell _{1}=J-1,\;\ell _{2}=J+1$ and $m_{S}$ is as defined in Eq. (42).
The system (52) reduses to a single equation for $J=0$ since $f^{J-1}(p)=0$
in that case.

Our equation (24), or its radial components (46), (49), (52), contain the
relativistic two-body kinematics (kinetic energy, recoil effects) exactly,
but the dynamics are included approximately due to the limited nature of our
trial state (20). This limitation is reflected in the fact that the
interaction kernels of our equations contain only ``tree-level'' Feynman
diagrams. Nevertheless our equations (46), (49), (52) have no
negative-energy solutions, in contrast to the BS equation. They are
variationally derived, hence the energy eigenvalues obtained from them will
give meaningful values for any strength of the coupling.

To our knowledge, it is not possible to obtain analytic solutions of the
relativistic radial momentum-space equations (46), (49) and (52). Thus one
must resort to numerical or other approximation methods. Numerical solutions
of such equations are discussed, for example, in [10], while a variational
approximation has been employed in [5]. However, in this paper we will
concentrate on perturbative $O(\alpha ^{4})$ solutions, since it is
important to verify that our\ equations yield the correct fine structure for
systems like positronium.

Our equations will yield energies which are incomplete beyond $O(\alpha
^{4}) $, because our variational trial state (20), as mentioned, reflects
only ``tree-level'' Feynman diagrams, that is no radiative corrections are
incorporated. One could, of course, augment them by the addition of
invariant matrix elements corresponding to higher-order Feynman diagrams
(including radiative corrections) to the existing $\mathcal{M}$-matrices in
the kernels of our equations, as is done in the BS formalism. Indeed, such
an approach has been used in a similar, though not variational, treatment of
positronium by Zhang and Koniuk [15]. These authors show that the inclusion
of invariant matrix elements corresponding to single-loop diagrams yields
positronium energy eigenstates which are accurate to $O\left( \alpha
^{5},\alpha ^{5}\ln \alpha \right) $. However such augmentation of the
kernels ``by hand'' would be contrary to the spirit of the present
variational treatment, and we shall not pursue it in this work.

{\normalsize 
}

{\normalsize \vskip 0.8truecm \noindent{\textbf{\large 6. Semi-relativistic
expansions and the non-relativistic limit}} }

{\normalsize \vskip 0.4truecm }

For perturbative solutions of our radial equations, it is necessary to work
out expansions of the relevant expressions to first order beyond the
non-relativistic limit. This shall be summarized in the present section. We
perform the calculation in the Coulomb gauge, in which the photon propagator
has the form [16] 
\begin{equation}
D_{00}\left( \mathbf{k}\right) =\frac{1}{\mathbf{k}^{2}},\;\;D_{0j}\left( 
\mathbf{k}\right) =0,\;\;D_{ij}\left( k^{\mu }\right) =\frac{1}{k^{\mu
}k_{\mu }}\left( \delta _{ij}-\frac{k_{i}k_{j}}{\mathbf{k}^{2}}\right) ,
\end{equation}
where $k^{\mu }=\left( \omega _{p}-\omega _{q},\mathbf{p-q}\right) $.

To expand the amplitudes $\mathcal{M}$\ of (26) and (27)\ to one order of $%
\left( p/m\right) ^{2}$ beyond the non-relativistic limit, we take the
free-particle spinors to be 
\begin{equation}
u(\mathbf{p,}i)=\left[ 
\begin{array}{c}
\left( 1+\frac{\mathbf{p}^{2}}{8m^{2}}\right) \\ 
\frac{(\overrightarrow{\sigma }\cdot \mathbf{p})}{2m}
\end{array}
\right] \varphi _{i},\;\ \ \ \ \ \ \ v(\mathbf{p,}i)=\left[ 
\begin{array}{c}
\frac{(\overrightarrow{\sigma }\mathbf{\cdot p})}{2m} \\ 
\left( 1+\frac{\mathbf{p}^{2}}{8m^{2}}\right)
\end{array}
\right] \chi _{i},
\end{equation}
as discussed in Appendix C. In this approximation the photon propagator
takes on the form

\begin{equation}
D_{00}\left( \mathbf{p-q}\right) =\frac{1}{\left( \mathbf{p-q}\right) ^{2}}%
,\;\;\;D_{ij}\left( \mathbf{p-q}\right) \simeq -\frac{1}{\left( \mathbf{p-q}%
\right) ^{2}}\left( \delta _{ij}-\frac{\left( p-q\right) _{i}\left(
p-q\right) _{j}}{\left( \mathbf{p-q}\right) ^{2}}\right) .
\end{equation}
Corresponding calculations give for the orbital part of the $\mathcal{M}$%
-matrix 
\begin{equation}
\mathcal{M}_{s_{1}s_{2}\sigma _{1}\sigma _{2}}^{ope\left( orb\right) }(%
\mathbf{p},\mathbf{q})=ie^{2}\left\{ \frac{1}{\left( \mathbf{p-q}\right) ^{2}%
}+\frac{1}{m^{2}}\left( \frac{1}{4}+\frac{\mathbf{q\cdot p}}{\left( \mathbf{%
p-q}\right) ^{2}}+\frac{\left( \mathbf{p}\times \mathbf{q}\right) ^{2}}{%
\left( \mathbf{p}-\mathbf{q}\right) ^{4}}\right) \right\} \delta
_{s_{1}\sigma _{1}}\delta _{s_{2}\sigma _{2}}.
\end{equation}
The terms of the expansion linear in spin correspond to the spin-orbit
interaction: 
\begin{equation}
\mathcal{M}_{s_{1}s_{2}\sigma _{1}\sigma _{2}}^{ope\left( s-o\right) }(%
\mathbf{p},\mathbf{q})=\frac{3e^{2}}{4m^{2}}\varphi _{s_{1}}^{\dagger }\chi
_{\sigma _{2}}^{\dagger }\frac{\left( \overrightarrow{\sigma }^{\left(
+\right) }-\overrightarrow{\sigma }^{\left( -\right) }\right) \cdot \left( 
\mathbf{p\times q}\right) }{\left( \mathbf{p-q}\right) ^{2}}\varphi _{\sigma
_{1}}\chi _{s_{2}}.
\end{equation}
Here $\overrightarrow{\sigma }^{\left( +\right) }$ and $\overrightarrow{%
\sigma }^{\left( -\right) }$\ are positron and electron\ spin matrices
respectively, defined as follow: $\overrightarrow{\sigma }^{\left( +\right)
}\varphi _{\sigma _{1}}\chi _{s_{2}}=\left( \overrightarrow{\sigma }^{\left(
+\right) }\varphi _{\sigma _{1}}\right) \chi _{s_{2}}$, $\overrightarrow{%
\sigma }^{\left( -\right) }\varphi _{\sigma _{1}}\chi _{s_{2}}=\varphi
_{\sigma _{1}}\left( \overrightarrow{\sigma }^{\left( -\right) }\chi
_{s_{2}}\right) $. The quadratic spin terms or spin-spin interaction terms
are 
\begin{eqnarray}
&&\mathcal{M}_{s_{1}s_{2}\sigma _{1}\sigma _{2}}^{ope\left( s-s\right) }(%
\mathbf{p},\mathbf{q})  \notag \\
&=&\frac{ie^{2}}{4m^{2}}\varphi _{s_{1}}^{\dagger }\chi _{\sigma
_{2}}^{\dagger }\left\{ -\frac{\left( \overrightarrow{\sigma }^{\left(
+\right) }\cdot \left( \mathbf{p-q}\right) \right) \left( \overrightarrow{%
\sigma }^{\left( -\right) }\cdot \left( \mathbf{p-q}\right) \right) }{\left( 
\mathbf{p-q}\right) ^{2}}+\overrightarrow{\sigma }^{\left( +\right) }\cdot 
\overrightarrow{\sigma }^{\left( -\right) }\right\} \varphi _{\sigma
_{1}}\chi _{s_{2}}.
\end{eqnarray}
Lastly, the virtual annihilation contribution is given by

{\normalsize 
\begin{equation}
\mathcal{M}_{s_{1}s_{2}\sigma _{1}\sigma _{2}}^{ann}(p,q)=-\frac{ie^{2}}{%
4m^{2}}\varphi _{s_{1}}^{\dagger }\chi _{\sigma _{2}}^{\dagger }\left\{ 
\overrightarrow{\sigma }^{\left( +\right) }\cdot \overrightarrow{\sigma }%
^{\left( -\right) }\right\} \varphi _{\sigma _{1}}\chi _{s_{2}},
\end{equation}
}where we have excluded a divergent term, which appears in the Coulomb gauge
calculation. This divergence is an artefact of the Coulomb gauge. It does
not arise, for example, in the Lorentz gauge, where only the expression (63)
is obtained. However the Lorentz gauge is not convenient for obtaining all
other $O(\alpha ^{4})$ corrections because it contains spurious degrees of
freedom (longitudinal polarization) of the photon.

We have used expressions (60)-(63) to obtain the corresponding radial
kernels. Details of the calculations can be found in Appendix D. We use the\
notation $z=\left( p^{2}+q^{2}\right) /2pq$, and\ $Q_{\lambda }(z)$ \ is the
Legendre function of the second kind [17]. The contributions of the various
terms to the kernel are as follows:

\noindent \textbf{Singlet states} with $\ell =J\ (J\geq
0),\;P=(-1)^{J+1},\;C=\left( -1\right) ^{J}$

Orbital term{\normalsize \ 
\begin{eqnarray}
&&\mathcal{K}^{\left( sgl\right) \left( o\right) }\left( p,q\right)  \notag
\\
&=&\frac{2\pi e^{2}}{pq}Q_{J}(z)+\frac{\pi e^{2}}{m^{2}}\left( -\frac{J-3}{2}%
\left( \frac{p}{q}+\frac{q}{p}\right) Q_{J}(z)+\left( J+1\right)
Q_{J+1}(z)-2\delta _{J,0}\right) .
\end{eqnarray}
}

Spin-orbit interaction\ {\normalsize 
\begin{equation}
\mathcal{K}^{\left( sgl\right) \left( s-o\right) }\left( p,q\right) =0.
\end{equation}
}

Spin-spin interaction\ {\normalsize 
\begin{equation}
\mathcal{K}^{\left( sgl\right) \left( s-s\right) }\left( p,q\right) =\frac{%
2\pi e^{2}}{m^{2}}\delta _{J,0}.
\end{equation}
}

{\normalsize \noindent \textbf{Triplet states }}with{\normalsize \textbf{\ } 
$\ell =J\ (J\geq 1),\;P=(-1)^{J+1},\;C=\left( -1\right) ^{J+1}$}

Orbital term {\normalsize 
\begin{equation}
K^{\left( tr\right) \left( o\right) }(p,q)=\frac{2\pi e^{2}}{pq}Q_{J}\left(
z\right) +\frac{\pi e^{2}}{m^{2}}\left( -\frac{J-3}{2}\left( \frac{p}{q}+%
\frac{q}{p}\right) Q_{J}\left( z\right) +\left( J+1\right) Q_{J+1}\left(
z\right) \right) .
\end{equation}
}

Spin-orbit interaction {\normalsize 
\begin{equation}
K^{\left( tr\right) \left( s-o\right) }(p,q)=-\frac{3\pi e^{2}}{m^{2}}\frac{1%
}{2J+1}\left\{ Q_{J+1}\left( z\right) -Q_{J-1}\left( z\right) \right\} .
\end{equation}
}

Spin-spin interaction {\normalsize 
\begin{eqnarray}
&&\mathcal{K}^{\left( tr\right) \left( s-s\right) }(p,q)  \notag \\
&=&\frac{\pi e^{2}}{2m^{2}}\left( \frac{p}{q}+\frac{q}{p}\right) Q_{J}\left(
z\right) -\frac{\pi e^{2}}{m^{2}}\frac{1}{2J+1}\left\{ JQ_{J+1}\left(
z\right) +\left( J+1\right) Q_{J-1}\left( z\right) \right\} .
\end{eqnarray}
}

{\normalsize \noindent }\textbf{Triplet states} with $\ell =J-1\ (J\geq
1),\;\ell =J+1\ (J\geq 0),\;P=(-1)^{J},\;C=\left( -1\right) ^{J}$

The off-diagonal elements of the kernel matrix (Eqs.. (52)-(54)), $K_{12}$
and $K_{21}$\thinspace\ which are responsible for mixing of states with $%
\ell =J-1$ and $\ell =J+1$, get a non-zero contribution from the spin-spin
interactions only:{\normalsize \ 
\begin{equation}
\mathcal{K}_{12}\left( p,q\right) =\mathcal{K}_{21}\left( p,q\right) =\frac{%
\pi e^{2}}{5m^{2}}\frac{\sqrt{J\left( J+1\right) }}{\left( 2J+1\right) }%
\left( \frac{p}{q}Q_{J+1}\left( z\right) +\frac{q}{p}Q_{J-1}\left( z\right)
-2Q_{J}\left( z\right) \right) .
\end{equation}
}The contributions to the diagonal elements of the kernel matrix are the
following:{\normalsize \ }

Orbital terms{\normalsize \ 
\begin{eqnarray}
\mathcal{K}_{11}^{\left( o\right) }\left( p,q\right)  &=&\frac{2\pi e^{2}}{pq%
}Q_{J-1}\left( z\right)   \notag \\
&&+\frac{\pi e^{2}}{m^{2}}\left( -\frac{J-4}{2}\left( \frac{p}{q}+\frac{q}{p}%
\right) Q_{J-1}\left( z\right) +JQ_{J}\left( z\right) -2\delta
_{J-1,0}\right) ,
\end{eqnarray}
\begin{eqnarray}
\mathcal{K}_{22}^{\left( o\right) }\left( p,q\right)  &=&\frac{2\pi e^{2}}{pq%
}Q_{J+1}\left( z\right)   \notag \\
&&+\frac{\pi e^{2}}{m^{2}}\left( -\frac{J-2}{2}\left( \frac{p}{q}+\frac{q}{p}%
\right) Q_{J+1}+\left( J+2\right) Q_{J+2}\right) .
\end{eqnarray}
}

Spin-orbit interaction{\normalsize \ 
\begin{equation}
\mathcal{K}_{11}^{\left( s-o\right) }\left( p,q\right) =\frac{3\pi e^{2}}{%
m^{2}}\frac{J-1}{2J-1}\left( Q_{J}\left( z\right) -Q_{J-2}\left( z\right)
\right) ,
\end{equation}
\begin{equation}
\mathcal{K}_{22}^{\left( s-o\right) }\left( p,q\right) =-\frac{3\pi e^{2}}{%
m^{2}c^{2}}\frac{J+2}{2J+3}\left( Q_{J+2}\left( z\right) -Q_{J}\left(
z\right) \right) .
\end{equation}
}

Spin-spin interaction{\normalsize \ 
\begin{equation}
\mathcal{K}_{11}^{\left( s-s\right) }\left( p,q\right) =\frac{\pi e^{2}}{%
2m^{2}}\frac{1}{2J+1}\left( \left( \frac{p}{q}+\frac{q}{p}\right)
Q_{J-1}\left( z\right) -2Q_{J}\left( z\right) \right) ,
\end{equation}
\begin{equation}
\mathcal{K}_{22}^{\left( s-s\right) }\left( p,q\right) =\frac{\pi e^{2}}{%
2m^{2}}\frac{1}{2J+3}\left( \left( \frac{p}{q}+\frac{q}{p}\right)
Q_{J+1}\left( z\right) -2Q_{J+2}\left( z\right) \right) .
\end{equation}
}

Annihilation term{\normalsize \ 
\begin{equation}
\mathcal{K}^{ann}\left( p,q\right) =-\frac{2\pi e^{2}}{m^{2}}\delta _{J-1,0}.
\end{equation}
}

We note that in the non-relativistic limit only the first terms of the
orbital part of the kernels survive. They have the common form $2\pi
ie^{2}Q_{\ell }(z)/pq$, hence all radial equations reduce to the form 
\begin{equation}
\left( 2\omega _{p}-E\right) f^{\ell }(p)=\frac{m^{2}e^{2}}{\pi \omega _{p}p}%
\int_{0}^{\infty }dq\frac{q}{\omega _{q}}Q_{\ell }(z)f^{\ell }(q).
\end{equation}
Recalling, also, that 
\begin{equation}
\omega _{p}=\sqrt{m^{2}+\mathbf{p}^{2}}\simeq m\left( 1+\frac{1}{2}\left( 
\frac{\mathbf{p}}{m}\right) ^{2}\right) ,
\end{equation}
we obtain, in the non relativistic limit, the momentum-space Schr\"{o}dinger
radial equations 
\begin{equation}
\left( \frac{\mathbf{p}^{2}}{2\mu }-\varepsilon \right) f^{\ell }(p)=\frac{%
\alpha }{\pi }\frac{1}{p}\int_{0}^{\infty }dq\,q\,Q_{J}(z)f^{\ell }(q),
\end{equation}
where $\alpha =e^{2}/4\pi $, $\mu =\frac{m}{2}$, $\varepsilon =E-2m$.

{\normalsize 
}

{\normalsize \vskip0.8truecm \noindent {\textbf{\large 7. Energy levels.
Fine and hyperfine structure}} }

{\normalsize \vskip 0.4truecm }

{\normalsize 
}

The relativistic energy eigenvalues $E_{n,J}$ can be calculated from the
expression 
\begin{eqnarray}
E\int_{0}^{\infty }dp\,p^{2}f^{J}(p)f^{J}(p) &=&\int_{0}^{\infty
}dp\,p^{2}\,2\omega _{p}f^{J}(p)f^{J}(p)  \notag \\
&&-\frac{m^{2}}{\left( 2\pi \right) ^{3}}\int_{0}^{\infty }\frac{dpp^{2}}{%
\omega _{p}}\int_{0}^{\infty }dq\frac{q^{2}}{\omega _{q}}\mathcal{K}^{\left(
sgl,tr\right) }(p\mathbf{,}q)f^{J}(p)f^{J}(q)
\end{eqnarray}
for the singlet and $\ell =J$ triplet states.

For the $\ell =J\mp 1$ triplet states the corresponding result is (see
equation (52))

{\normalsize 
\begin{eqnarray}
E\int_{0}^{\infty }dp\,p^{2}\mathbb{F}^{\dagger }(p)\mathbb{F}(p)
&=&\int_{0}^{\infty }dp\,p^{2}\,2\omega _{p}\mathbb{F}^{\dagger }(p)\mathbb{F%
}(p)  \notag \\
&&-\frac{m^{2}}{\left( 2\pi \right) ^{3}}\int_{0}^{\infty }\frac{dpp^{2}}{%
\omega _{p}}\int_{0}^{\infty }dq\frac{q^{2}}{\omega _{q}}\mathbb{K}(p\mathbf{%
,}q)\mathbb{F}^{\dagger }(p)\mathbb{F}(q).
\end{eqnarray}
}

To obtain results for $E$ to $O\left( \alpha ^{4}\right) $ we use the forms
of the kernels expanded to $O\left( p^{2}/m^{2}\right) $ (eqs. (64)-(77))
and replace $f^{\ell }(p)$ by their non-relativistic (Schr\"{o}dinger) form
(see (147), Appendix D). The most important integrals that we used for
calculating (81) and (82), are given in Appendix D. In Appendix E we show
that the contribution of kernels $K_{12}$ and $K_{21}$ in (82), is zero at $%
O(\alpha ^{4})$. Thus, the energy corrections for the triplet states with $%
\ell =J-1$ and $\ell =J+1$ can be calculated independently.

\noindent The results will be presented in the form $\Delta \varepsilon
=E-2m+\alpha ^{2}m/2n^{2}$.

\textbf{Singlet states} $\left( \ell =J\ (J\geq 0),\;P=(-1)^{J+1},\;C=\left(
-1\right) ^{J}\right) $

\noindent The kinetic energy corrections 
\begin{equation}
\Delta \varepsilon _{K}^{(sgl)}=-\frac{\alpha ^{4}m}{8}\left( \frac{1}{2J+1}%
\frac{1}{n^{3}}-\frac{3}{8}\frac{1}{n^{4}}\right) .
\end{equation}
The potential energy corrections {\normalsize 
\begin{equation}
\Delta \varepsilon _{P}^{(sgl)\left( o\right) }=-\frac{\alpha ^{4}m}{8}%
\left( \left( \frac{3}{2J+1}-2\delta _{J,0}\right) \frac{1}{n^{3}}-\frac{1}{%
n^{4}}\right) ,
\end{equation}
}

{\normalsize 
\begin{equation}
\Delta \varepsilon _{P}^{(sgl)\left( s-o\right) }=0,
\end{equation}
}

\begin{equation}
\Delta \varepsilon _{P}^{(sgl)\left( s-s\right) }=-\frac{\alpha ^{4}m}{4}%
\frac{\delta _{J,0}}{n^{3}}.
\end{equation}
The total energy corrections 
\begin{equation}
\Delta \varepsilon ^{(sgl)}=-\frac{\alpha ^{4}m}{8}\left( \frac{4}{2J+1}%
\frac{1}{n^{3}}-\frac{11}{8n^{4}}\right) .
\end{equation}

\textbf{Triplet states}{\normalsize \ $\left( \ell =J\ (J\geq
1),\;P=(-1)^{J+1},\;C=\left( -1\right) ^{J+1}\right) $}

{\normalsize \noindent }The kinetic energy corrections 
\begin{equation}
\Delta \varepsilon _{K}^{(tr)}=-\frac{\alpha ^{4}m}{8}\left( \frac{1}{2J+1}%
\frac{1}{n^{3}}-\frac{3}{8}\frac{1}{n^{4}}\right) .
\end{equation}
The potential energy corrections 
\begin{equation}
\Delta \varepsilon _{P}^{\left( tr\right) \left( o\right) }=-\frac{\alpha
^{4}m}{8}\left( \left( \frac{3}{2J+1}-2\delta _{J,0}\right) \frac{1}{n^{3}}-%
\frac{1}{n^{4}}\right) ,
\end{equation}
\begin{equation}
\Delta \varepsilon _{P}^{\left( tr\right) \left( s-o\right) }=-\frac{\alpha
^{4}m}{8}\frac{3}{J(J+1)\left( 2J+1\right) }\frac{1}{n^{3}},
\end{equation}
\begin{equation}
\Delta \varepsilon _{P}^{\left( s-s\right) }=\frac{\alpha ^{4}m}{8}\frac{1}{%
J\left( J+1\right) \left( 2J+1\right) }\frac{1}{n^{3}}.
\end{equation}
The total energy corrections {\normalsize 
\begin{equation}
\Delta \varepsilon ^{(tr)}=-\frac{\alpha ^{4}m}{8}\left( \left( \frac{4}{2J+1%
}+\frac{2}{J\left( J+1\right) \left( 2J+1\right) }\right) \frac{1}{n^{3}}-%
\frac{11}{8}\frac{1}{n^{4}}\right) .
\end{equation}
}

\textbf{Triplet states}{\normalsize \ $\left( \ell =J-1\ (J\geq
1),\;\;P=(-1)^{J},\;C=\left( -1\right) ^{J}\right) $}

{\normalsize \noindent }The kinetic energy corrections 
\begin{equation}
\Delta \varepsilon _{K}^{(tr)(J-1)}=-\frac{\alpha ^{4}m}{8}\left( \frac{1}{%
2J-1}\frac{1}{n^{3}}-\frac{3}{8}\frac{1}{n^{4}}\right) .
\end{equation}
The potential energy corrections{\normalsize \ 
\begin{equation}
\Delta \varepsilon _{P}^{\left( tr\right) \left( o\right) (J-1)}=-\frac{%
\alpha ^{4}m}{8}\left[ \left( \frac{3}{2J-1}-2\delta _{J,1}\right) \frac{1}{%
n^{3}}-\frac{1}{n^{4}}\right] ,
\end{equation}
\begin{equation}
\Delta \varepsilon _{P}^{tr\left( s-o\right) (J-1)}=\frac{\alpha ^{4}m}{8}%
\frac{3\left( 1-\delta _{J,1}\right) }{J\left( 2J-1\right) }\frac{1}{n^{3}},
\end{equation}
\begin{equation}
\Delta \varepsilon _{P}^{tr\left( s-s\right) (J-1)}=-\frac{\alpha ^{4}m}{8}%
\frac{1-\delta _{J,1}}{J\left( 2J+1\right) \left( 2J-1\right) }\frac{1}{n^{3}%
},
\end{equation}
}

{\normalsize 
\begin{equation}
\Delta \varepsilon ^{\left( ann\right) }=\frac{\alpha ^{4}m}{4}\frac{1}{n^{3}%
}\delta _{J,1}.
\end{equation}
}

{\normalsize \noindent }The total energy corrections{\normalsize 
\begin{equation}
\Delta \varepsilon ^{tr(J-1)}=-\frac{\alpha ^{4}m}{8}\left( \left( \frac{4}{%
2J-1}-\frac{2\left( 3J+1\right) \left( 1-\delta _{J,1}\right) }{J\left(
2J+1\right) \left( 2J-1\right) }-4\delta _{J,1}\right) \frac{1}{n^{3}}-\frac{%
11}{8}\frac{1}{n^{4}}\right) .
\end{equation}
}

\textbf{Triplet states}{\normalsize \textbf{\ }$\left( \ell =J+1\ (J\geq
0),\;\;P=(-1)^{J},\;C=\left( -1\right) ^{J}\right) $}

{\normalsize \noindent }The kinetic energy corrections 
\begin{equation}
\Delta \varepsilon _{K}^{(tr)(J+1)}=-\frac{\alpha ^{4}m}{8}\left( \frac{1}{%
2J+3}\frac{1}{n^{3}}-\frac{3}{8}\frac{1}{n^{4}}\right) .
\end{equation}
The potential energy corrections {\normalsize 
\begin{equation}
\Delta \varepsilon _{P}^{\left( tr\right) \left( o\right) (J+1)}=-\frac{%
\alpha ^{4}m}{8}\left[ \frac{3}{2J+3}\frac{1}{n^{3}}-\frac{1}{n^{4}}\right] ,
\end{equation}
\begin{equation}
\Delta \varepsilon _{P}^{tr\left( s-o\right) (J+1)}=-\frac{\alpha ^{4}m}{8}%
\frac{3}{\left( J+1\right) \left( 2J+3\right) }\frac{1}{n^{3}},
\end{equation}
\begin{equation}
\Delta \varepsilon _{P}^{tr\left( s-s\right) (J+1)}=-\frac{\alpha ^{4}m}{8}%
\frac{1}{\left( J+1\right) \left( 2J+3\right) \left( 2J+1\right) }\frac{1}{%
n^{3}}.
\end{equation}
}The total energy corrections 
\begin{equation}
\Delta \varepsilon ^{tr(J+1)}=-\frac{\alpha ^{4}m}{8}\left( \frac{2}{2J+3}%
\left( 2+\frac{3J+2}{\left( J+1\right) \left( 2J+1\right) }\right) \frac{1}{%
n^{3}}-\frac{11}{8}\frac{1}{n^{4}}\right) .
\end{equation}

These results are in agreement with the well-known positronium fine
structure results [18], [19].

{\normalsize \vskip 0.8truecm \noindent{\textbf{\large 8. Concluding remarks}%
} }

{\normalsize 
}

{\normalsize \vskip 0.4truecm }

{\normalsize 
}We have considered a reformulation of electrodynamics, in which covariant
Green functions are used to solve the field equations for the mediating
electromagnetic field in terms of the fermion field. This leads to a
reformulated Hamiltonian with an interaction term in which the photon
propagator appears sandwiched between fermionic currents.

The variational method within a Hamiltonian formalism of quantum field
theory is used to determine approximate eigensolutions for bound
relativistic fermion-antifermion states. The reformulation enables us to use
the simplest possible trial state to derive a relativistic momentum-space
Salpeter-like equation for a positronium-like system. The invariant $%
\mathcal{M}$ matrices corresponding to one-photon exchange and virtual
annihilation Feynman diagrams arise directly in the interaction kernel of
this equation.

The trial states are chosen to be eigenstates of the total angular momentum
operator $\widehat{\mathbf{J}}^{2}$ and $\widehat{J}_{3}$, along with parity
and charge conjugation. A general relativistic reduction of the wave
equations to radial form is given. For given $J$ there is a single radial
equation for total spin zero singlet states, but for spin triplet states
there are, in general two coupled equations. We show how the classification
of states follows naturally from the system of eigenvalue equations obtained
with our trial state.

It is not possible, as far as we know, to obtain analytic solutions of our
relativistic radial equations nor the resulting eigenvalues of the
particle-antiparticle system described. However, it is possible to obtain $%
O(\alpha ^{4})$ corrections analytically for all states using perturbation
theory. The results agree with well known results for positronium, obtained
on the basis of the Bethe-Salpeter equation [19], which lends credence to
the validity of our variationally derived equations.

The method presented here can be generalized to include effects higher order
in alpha by using dressed propagators in place of the bare propagators. This
shall be the subject of a forthcoming work.

{\normalsize \vskip 0.8truecm \noindent{\textbf{\large Acknowledgments}} } 
{\normalsize \vskip 0.4truecm }

The financial support of the Natural Sciences and Engineering Research
Council of Canada for this work is gratefully acknowledged.

{\normalsize 


{\normalsize \vskip 0.8truecm \noindent{\textbf{\large Appendix A. Total angular momentum operator in relativistic form}}}

{\normalsize \vskip 0.4truecm }

{\normalsize 
}

The total angular momentum operator is defined by expression{\normalsize \ 
\begin{equation}
\widehat{\mathbf{J}}=\int d^{3}\mathbf{x}:\psi ^{\dagger }\left( x\right) 
\mathbf{(}\widehat{\mathbf{L}}+\widehat{\mathbf{S}})\psi \left( x\right) :,
\end{equation}
}where $\widehat{\mathbf{L}}=\widehat{\mathbf{x}}\times \widehat{\mathbf{p}}$
and $\widehat{\mathbf{S}}=\frac{1}{2}\widehat{\overrightarrow{\sigma }}$ are
the orbital angular momentum and spin operators. We use the standard
representation for the Pauli matrices{\normalsize \ 
\begin{equation}
\widehat{\overrightarrow{\sigma }}=\left[ 
\begin{array}{cc}
\overrightarrow{\sigma } & 0 \\ 
0 & \overrightarrow{\sigma }
\end{array}
\right] ,
\end{equation}
\begin{equation}
\sigma _{1}=\left[ 
\begin{array}{cc}
0 & 1 \\ 
1 & 0
\end{array}
\right] ,\;\;\ \ \ \sigma _{2}=\left[ 
\begin{array}{cc}
0 & -i \\ 
i & 0
\end{array}
\right] ,\;\;\;\sigma _{3}=\left[ 
\begin{array}{cc}
1 & 0 \\ 
0 & -1
\end{array}
\right] .
\end{equation}
}Using the field operator $\psi \left( x\right) $ in the form (15), after
tedious calculations we obtain 
\begin{equation*}
\widehat{J}_{1}=\int d^{3}\mathbf{q}\left( 
\begin{array}{c}
\widehat{L}_{q1}\left( b_{\mathbf{q}\uparrow }^{\dagger }b_{\mathbf{q}%
\uparrow }+b_{\mathbf{q}\downarrow }^{\dagger }b_{\mathbf{q}\downarrow }+d_{%
\mathbf{q}\uparrow }^{\dagger }d_{\mathbf{q}\uparrow }+d_{\mathbf{q}%
\downarrow }^{\dagger }d_{\mathbf{q}\downarrow }\right) \\ 
+\frac{1}{2}\left( b_{\mathbf{q}\uparrow }^{\dagger }b_{\mathbf{q}\downarrow
}+b_{\mathbf{q}\downarrow }^{\dagger }b_{\mathbf{q}\uparrow }+d_{\mathbf{q}%
\downarrow }^{\dagger }d_{\mathbf{q}\uparrow }+d_{\mathbf{q}\uparrow
}^{\dagger }d_{\mathbf{q}\downarrow }\right)
\end{array}
\right) ,
\end{equation*}
\begin{equation}
\widehat{J}_{2}=\int d^{3}\mathbf{q}\left( 
\begin{array}{c}
\widehat{L}_{q2}\left( b_{\mathbf{q}\uparrow }^{\dagger }b_{\mathbf{q}%
\uparrow }+b_{\mathbf{q}\downarrow }^{\dagger }b_{\mathbf{q}\downarrow }+d_{%
\mathbf{q}\uparrow }^{\dagger }d_{\mathbf{q}\uparrow }+d_{\mathbf{q}%
\downarrow }^{\dagger }d_{\mathbf{q}\downarrow }\right) \\ 
+\frac{i}{2}\left( -b_{\mathbf{q}\uparrow }^{\dagger }b_{\mathbf{q}%
\downarrow }+b_{\mathbf{q}\downarrow }^{\dagger }b_{\mathbf{q}\uparrow }-d_{%
\mathbf{q}\uparrow }^{\dagger }d_{\mathbf{q}\downarrow }+d_{\mathbf{q}%
\downarrow }^{\dagger }d_{\mathbf{q}\uparrow }\right)
\end{array}
\right) ,
\end{equation}
\begin{equation*}
\widehat{J}_{3}=\int d^{3}\mathbf{q}\left( 
\begin{array}{c}
\widehat{L}_{q3}\left( b_{\mathbf{q}\uparrow }^{\dagger }b_{\mathbf{q}%
\uparrow }+b_{\mathbf{q}\downarrow }^{\dagger }b_{\mathbf{q}\downarrow }+d_{%
\mathbf{q}\uparrow }^{\dagger }d_{\mathbf{q}\uparrow }+d_{\mathbf{q}%
\downarrow }^{\dagger }d_{\mathbf{q}\downarrow }\right) \\ 
+\frac{1}{2}\left( b_{\mathbf{q}\uparrow }^{\dagger }b_{\mathbf{q}\uparrow
}-b_{\mathbf{q}\downarrow }^{\dagger }b_{\mathbf{q}\downarrow }+d_{\mathbf{q}%
\uparrow }^{\dagger }d_{\mathbf{q}\uparrow }-d_{\mathbf{q}\downarrow
}^{\dagger }d_{\mathbf{q}\downarrow }\right)
\end{array}
\right) .
\end{equation*}
Here $\widehat{\mathbf{L}}_{q}$ is the orbital angular momentum operator in
momentum representation:

{\normalsize 
\begin{equation}
(\widehat{\mathbf{L}}_{q})_{i}\equiv \widehat{L}_{qi}=-i\left( \mathbf{%
q\times }\nabla _{q}\right) _{i}.
\end{equation}
}Note that these expressions are valid for any $t$, since the time-dependent
phase factors of the form $e^{i\omega _{q}t}$\ cancel out.

For the operator $\widehat{\mathbf{J}}^{2}=\widehat{J}_{1}^{2}+\widehat{J}%
_{2}^{2}+\widehat{J}_{3}^{2}$ we have{\normalsize \ 
\begin{equation*}
\widehat{\mathbf{J}}^{2}=\int d^{3}\mathbf{q}\left( 
\begin{array}{c}
\left( \widehat{\mathbf{L}}_{q}^{2}+\frac{3}{4}\right) \left( b_{\mathbf{q}%
\uparrow }^{\dagger }b_{\mathbf{q}\uparrow }+b_{\mathbf{q}\downarrow
}^{\dagger }b_{\mathbf{q}\downarrow }+d_{\mathbf{q}\uparrow }^{\dagger }d_{%
\mathbf{q}\uparrow }+d_{\mathbf{q}\downarrow }^{\dagger }d_{\mathbf{q}%
\downarrow }\right) \\ 
+\widehat{L}_{q-}b_{\mathbf{q}\uparrow }^{\dagger }b_{\mathbf{q}\downarrow }+%
\overset{\symbol{94}}{L}_{q+}b_{\mathbf{q}\downarrow }^{\dagger }b_{\mathbf{q%
}\uparrow }+\overset{\symbol{94}}{L}_{q-}d_{\mathbf{q}\uparrow }^{\dagger
}d_{\mathbf{q}\downarrow }+\overset{\symbol{94}}{L}_{q+}d_{\mathbf{q}%
\downarrow }^{\dagger }d_{\mathbf{q}\uparrow } \\ 
+\widehat{L}_{q3}\left( b_{\mathbf{q}\uparrow }^{\dagger }b_{\mathbf{q}%
\uparrow }-b_{\mathbf{q}\downarrow }^{\dagger }b_{\mathbf{q}\downarrow }+d_{%
\mathbf{q}\uparrow }^{\dagger }d_{\mathbf{q}\uparrow }-d_{\mathbf{q}%
\downarrow }^{\dagger }d_{\mathbf{q}\downarrow }\right)
\end{array}
\right) +
\end{equation*}
} 
\begin{equation}
+\frac{1}{2}\int d^{3}\mathbf{q}^{\prime }d^{3}\mathbf{q}\left( 
\begin{array}{c}
2\widehat{\mathbf{L}}_{q^{\prime }}\cdot \widehat{\mathbf{L}}_{q}\left( 
\begin{array}{c}
b_{\mathbf{q}^{\prime }\uparrow }^{\dagger }b_{\mathbf{q}^{\prime }\uparrow
}d_{\mathbf{q}\uparrow }^{\dagger }d_{\mathbf{q}\uparrow }+b_{\mathbf{q}%
^{\prime }\uparrow }^{\dagger }b_{\mathbf{q}^{\prime }\uparrow }d_{\mathbf{q}%
\downarrow }^{\dagger }d_{\mathbf{q}\downarrow } \\ 
+b_{\mathbf{q}^{\prime }\downarrow }^{\dagger }b_{\mathbf{q}^{\prime
}\downarrow }d_{\mathbf{q}\uparrow }^{\dagger }d_{\mathbf{q}\uparrow }+b_{%
\mathbf{q}^{\prime }\downarrow }^{\dagger }b_{\mathbf{q}^{\prime }\downarrow
}d_{\mathbf{q}\downarrow }^{\dagger }d_{\mathbf{q}\downarrow }
\end{array}
\right) \\ 
+\frac{1}{2}\left( b_{q^{\prime }\uparrow }^{\dagger }b_{q^{\prime }\uparrow
}d_{q\uparrow }^{\dagger }d_{q\uparrow }-b_{q^{\prime }\uparrow }^{\dagger
}b_{q^{\prime }\uparrow }d_{q\downarrow }^{\dagger }d_{q\downarrow }\right)
\\ 
-\frac{1}{2}\left( b_{q^{\prime }\downarrow }^{\dagger }b_{q^{\prime
}\downarrow }d_{q\uparrow }^{\dagger }d_{q\uparrow }-b_{q^{\prime
}\downarrow }^{\dagger }b_{q^{\prime }\downarrow }d_{q\downarrow }^{\dagger
}d_{q\downarrow }\right) \\ 
+b_{q^{\prime }\uparrow }^{\dagger }b_{q^{\prime }\downarrow }d_{q\downarrow
}^{\dagger }d_{q\uparrow }+b_{q^{\prime }\downarrow }^{\dagger }b_{q^{\prime
}\uparrow }d_{q\uparrow }^{\dagger }d_{q\downarrow } \\ 
+\widehat{L}_{q^{\prime }+}\left( 
\begin{array}{c}
b_{\mathbf{q}^{\prime }\uparrow }^{\dagger }b_{\mathbf{q}^{\prime }\uparrow
}d_{\mathbf{q}\downarrow }^{\dagger }d_{\mathbf{q}\uparrow }+b_{\mathbf{q}%
^{\prime }\downarrow }^{\dagger }b_{\mathbf{q}^{\prime }\downarrow }d_{%
\mathbf{q}\downarrow }^{\dagger }d_{\mathbf{q}\uparrow } \\ 
+b_{\mathbf{q}\downarrow }^{\dagger }b_{\mathbf{q}\uparrow }d_{\mathbf{q}%
^{\prime }\uparrow }^{\dagger }d_{\mathbf{q}^{\prime }\uparrow }+b_{\mathbf{q%
}\downarrow }^{\dagger }b_{\mathbf{q}\uparrow }d_{\mathbf{q}^{\prime
}\downarrow }^{\dagger }d_{\mathbf{q}^{\prime }\downarrow }
\end{array}
\right) \\ 
+\widehat{L}_{q^{\prime }-}\left( 
\begin{array}{c}
b_{\mathbf{q}^{\prime }\uparrow }^{\dagger }b_{\mathbf{q}^{\prime }\uparrow
}d_{\mathbf{q}\uparrow }^{\dagger }d_{\mathbf{q}\downarrow }+b_{\mathbf{q}%
^{\prime }\downarrow }^{\dagger }b_{\mathbf{q}^{\prime }\downarrow }d_{%
\mathbf{q}\uparrow }^{\dagger }d_{\mathbf{q}\downarrow } \\ 
+b_{\mathbf{q}\uparrow }^{\dagger }b_{\mathbf{q}\downarrow }d_{\mathbf{q}%
^{\prime }\uparrow }^{\dagger }d_{\mathbf{q}^{\prime }\uparrow }+b_{\mathbf{q%
}\uparrow }^{\dagger }b_{\mathbf{q}\downarrow }d_{\mathbf{q}^{\prime
}\downarrow }^{\dagger }d_{\mathbf{q}^{\prime }\downarrow }
\end{array}
\right) \\ 
+\left( \widehat{L}_{q^{\prime }3}+\widehat{L}_{q3}\right) \left( b_{\mathbf{%
q}^{\prime }\uparrow }^{\dagger }b_{\mathbf{q}^{\prime }\uparrow }d_{\mathbf{%
q}\uparrow }^{\dagger }d_{\mathbf{q}\uparrow }-b_{\mathbf{q}^{\prime
}\downarrow }^{\dagger }b_{\mathbf{q}^{\prime }\downarrow }d_{\mathbf{q}%
\downarrow }^{\dagger }d_{\mathbf{q}\downarrow }\right) \\ 
-\left( \widehat{L}_{q^{\prime }3}-\widehat{L}_{q3}\right) \left( b_{\mathbf{%
q}^{\prime }\uparrow }^{\dagger }b_{\mathbf{q}^{\prime }\uparrow }d_{\mathbf{%
q}\downarrow }^{\dagger }d_{\mathbf{q}\downarrow }-b_{\mathbf{q}^{\prime
}\downarrow }^{\dagger }b_{\mathbf{q}^{\prime }\downarrow }d_{\mathbf{q}%
\uparrow }^{\dagger }d_{\mathbf{q}\uparrow }\right)
\end{array}
\right) ,
\end{equation}
where 
\begin{equation}
\widehat{L}_{q+}=\widehat{L}_{q1}+i\widehat{L}_{q2},\;\;\;\;\;\;\;\widehat{L}%
_{q-}=\widehat{L}_{q1}-i\widehat{L}_{q2}\,.
\end{equation}
The requirement that the trial state (20) be an eigenstate of $\widehat{%
\mathbf{J}}^{2}$ and $\widehat{J}_{z}$ leads to the system of equations 
{\normalsize 
\begin{eqnarray}
\left( \widehat{L}_{3}+1\right) F_{11} &=&m_{J}F_{11},  \notag \\
\widehat{L}_{3}F_{12} &=&m_{J}F_{12},  \notag \\
\widehat{L}_{3}F_{21} &=&m_{J}F_{21}, \\
\left( \widehat{L}_{3}-1\right) F_{22} &=&m_{J}F_{22},  \notag
\end{eqnarray}
} 
\begin{eqnarray}
\left( J(J+1)-\widehat{\mathbf{L}}^{2}-2-2\widehat{L}_{3}\right) F_{11} &=&%
\widehat{L}_{-}\left( F_{12}+F_{21}\right) ,  \notag \\
\left( J(J+1)-\widehat{\mathbf{L}}^{2}-1\right) F_{12} &=&F_{21}+\widehat{L}%
_{+}F_{11}+\widehat{L}_{-}F_{22},  \notag \\
\left( J(J+1)-\widehat{\mathbf{L}}^{2}-1\right) F_{21} &=&F_{12}+\widehat{L}%
_{+}F_{11}+\widehat{L}_{-}F_{22}, \\
\left( J(J+1)-\widehat{\mathbf{L}}^{2}-2+2\widehat{L}_{3}\right) F_{22} &=&%
\widehat{L}_{+}\left( F_{12}+F_{21}\right) .  \notag
\end{eqnarray}
Substitution of the expressions (33) for $F_{s_{1}s_{2}}$ and use of eq.
(111)\ gives{\normalsize \ 
\begin{equation}
m_{12}=m_{21}=m_{J},\;\ \ \ \ m_{11}=m_{J}-1,\;\ \ \ \ \ \ m_{22}=m_{J}+1,
\end{equation}
\begin{equation}
\ell _{11}=\ell _{22}=\ell _{12}=\ell _{21}=\ell ,
\end{equation}
}and{\normalsize \ 
\begin{eqnarray}
\left( J(J+1)-\ell (\ell +1)-2m_{J}\right) f_{11}^{\ell }(p) &=&\sqrt{(\ell
-m_{J}+1)(\ell +m_{J})}f_{12}^{\ell }(p)  \notag \\
&&+\sqrt{(\ell -m_{J}+1)(\ell +m_{J})}f_{21}^{\ell }(p),  \notag \\
\left( J(J+1)-\ell (\ell +1)-1\right) f_{12}^{\ell }(p) &=&f_{21}^{\ell }(p)
\notag \\
&&+\sqrt{(\ell +m_{J})(\ell -m_{J}+1)}f_{11}^{\ell }(p)  \notag \\
&&+\sqrt{(\ell -m_{J})(\ell +m_{J}+1)}f_{22}^{\ell }(p),  \notag \\
\left( J(J+1)-\ell (\ell +1)-1\right) f_{21}^{\ell }(p) &=&f_{12}^{\ell }(p)
\\
&&+\sqrt{(\ell +m_{J})(\ell -m_{J}+1)}f_{11}^{\ell }(p)  \notag \\
&&+\sqrt{(\ell -m_{J})(\ell +m_{J}+1)}f_{22}^{\ell }(p),  \notag \\
\left( J(J+1)-\ell (\ell +1)+2m_{J}\right) f_{22}^{\ell }(p) &=&\sqrt{(\ell
+m_{J}+1)(\ell -m_{J})}f_{12}^{\ell }(p)  \notag \\
&&+\sqrt{(\ell +m_{J}+1)(\ell -m_{J})}f_{21}^{\ell }(p).  \notag
\end{eqnarray}
}

The singlet states correspond to the solution $f_{11}^{\ell }\left( p\right)
=f_{22}^{\ell }\left( p\right) =0$, $f_{12}^{\ell }\left( p\right)
=-f_{21}^{\ell }\left( p\right) $\ of this system with $\ell =J$ $(J\geq 0)$.

For the triplet states the solutions are $f_{12}^{\ell }\left( p\right)
=f_{21}^{\ell }\left( p\right) \equiv f^{\ell }\left( p\right) $, and,

\noindent for \ $\ell =J-1$ $(J\geq 1)$: 
\begin{eqnarray}
\left( J-m_{J}\right) f_{11}^{J-1}(p) &=&\sqrt{\left( J-m_{J}\right)
(J+m_{J}-1)}f^{J-1}(p), \\
\left( J+m_{J}\right) f_{22}^{J-1}(p) &=&\sqrt{\left( J+m_{J}\right)
(J-m_{J}-1)}f^{J-1}(p),
\end{eqnarray}
for\ $\ell =J$ $(J\geq 1)$: 
\begin{eqnarray}
m_{J}f_{11}^{J}(p) &=&-\sqrt{\left( J+m_{J}\right) (J-m_{J}+1)}f^{J}(p), \\
m_{J}f_{22}^{J}(p) &=&\sqrt{\left( J-m_{J}\right) (J+m_{J}+1)}f^{J}(p),
\end{eqnarray}
for\ $\ell =J+1$ $(J\geq 0)$: 
\begin{eqnarray}
\left( J+1+m_{J}\right) f_{11}^{J+1}(p) &=&-\sqrt{\left( J-m_{J}+2\right)
\left( J+m_{J}+1\right) }f^{J+1}(p), \\
\left( J+1-m_{J}\right) f_{22}^{J+1}(p) &=&-\sqrt{\left( J-m_{J}+1\right)
\left( J+m_{J}+2\right) }f^{J+1}(p).
\end{eqnarray}

It is convenient to introduce the table of coefficients $C_{Jm_{J}}^{\left(
tr\right) \ell m_{s}}$ :

\begin{center}
\begin{tabular}{||c|c|c|c||}
\hline
& $m_{s}=+1$ & $m_{s}=0$ & $m_{s}=-1$ \\ \hline
$\ell =J-1$ & $\sqrt{\frac{\left( J+m_{J}-1\right) (J+m_{J})}{J\left(
2J-1\right) }}$ & $\sqrt{\frac{\left( J-m_{J}\right) \left( J+m_{J}\right) }{%
J\left( 2J-1\right) }}$ & $\sqrt{\frac{\left( J-m_{J}-1\right) (J-m_{J})}{%
J\left( 2J-1\right) }}$ \\ \hline
$\ell =J$ & $-\sqrt{\frac{\left( J+m_{J}\right) (J-m_{J}+1)}{J\left(
J+1\right) }}$ & $\frac{m_{J}}{\sqrt{J\left( J+1\right) }}$ & $\sqrt{\frac{%
\left( J-m_{J}\right) (J+m_{J}+1)}{J\left( J+1\right) }}$ \\ \hline
$\ell =J+1$ & $\sqrt{\frac{\left( J-m_{J}+1\right) \left( J-m_{J}+2\right) }{%
\left( J+1\right) \left( 2J+3\right) }}$ & $-\sqrt{\frac{\left(
J-m_{J}+1\right) \left( J+m_{J}+1\right) }{\left( J+1\right) \left(
2J+3\right) }}$ & $\sqrt{\frac{\left( J+m_{J}+2\right) \left(
J+m_{J}+1\right) }{\left( J+1\right) \left( 2J+3\right) }}$ \\ \hline
\end{tabular}
\end{center}

These coefficients coincide with the usual Clebsch-Gordan coefficients for $%
S=1$ except for a factor $2$ in the denominator, which we absorb into the
normalization constant.

{\normalsize 
}

{\normalsize \vskip 0.8truecm \noindent{\textbf{\large Appendix B. Parity
and charge conjugation}} }

{\normalsize \vskip 0.4truecm }

{\normalsize 
}

We consider the application of the parity operator to the trial state (20): 
\begin{eqnarray}
\widehat{\mathcal{P}}\left| \psi _{T}\right\rangle &=&\underset{s_{1}s_{2}}{%
\sum }\int d^{3}\mathbf{p}F_{s_{1}s_{2}}(\mathbf{p})\widehat{\mathcal{P}}%
b_{ps_{1}}^{\dagger }d_{-ps_{2}}^{\dagger }\left| 0\right\rangle  \notag \\
&=&\underset{s_{1}s_{2}}{\sum }\int d^{3}\mathbf{p}F_{s_{1}s_{2}}(\mathbf{p})%
\widehat{\mathcal{P}}b_{ps_{1}}^{\dagger }\widehat{\mathcal{P}}^{-1}\widehat{%
\mathcal{P}}d_{-ps_{2}}^{\dagger }\widehat{\mathcal{P}}^{-1}\widehat{%
\mathcal{P}}\left| 0\right\rangle .
\end{eqnarray}
Making use of the properties 
\begin{equation}
\widehat{\mathcal{P}}b_{ps_{1}}^{\dagger }\widehat{\mathcal{P}}^{-1}=\eta
^{P}b_{-ps_{1}}^{\dagger },\;\;\;\;\;\widehat{\mathcal{P}}%
d_{-ps_{2}}^{\dagger }\widehat{\mathcal{P}}^{-1}=-\eta
^{P}d_{ps_{2}}^{\dagger },\;\;\;\;\;\widehat{\mathcal{P}}\left|
0\right\rangle =\left| 0\right\rangle ,
\end{equation}
where $\eta ^{P}$ is the intrinsic parity ($\left( \eta ^{P}\right) ^{2}=1$%
), it follows that 
\begin{equation*}
\widehat{\mathcal{P}}\left| \psi _{T}\right\rangle =\underset{s_{1}s_{2}}{%
\sum }\int d^{3}\mathbf{p}F_{s_{1}s_{2}}(\mathbf{p})\widehat{\mathcal{P}}%
b_{ps_{1}}^{\dagger }d_{-ps_{2}}^{\dagger }\left| 0\right\rangle =\underset{%
s_{1}s_{2}}{-\sum }\int d^{3}\mathbf{p}F_{s_{1}s_{2}}(-\mathbf{p}%
)b_{ps_{1}}^{\dagger }d_{-ps_{2}}^{\dagger }\left| 0\right\rangle
\end{equation*}
\begin{equation}
=P\underset{s_{1}s_{2}}{\sum }\int d^{3}\mathbf{p}F_{s_{1}s_{2}}(\mathbf{p}%
)b_{ps_{1}}^{\dagger }d_{-ps_{2}}^{\dagger }\left| 0\right\rangle ,
\end{equation}
where the parity eigenvalue $P$ depends on the symmetry of $%
F_{s_{1}s_{2}}\left( p\right) $ in different states:

\noindent For the singlet states\ $\left( \ell =J\right) $ we get from (36) $%
F_{s_{1}s_{2}}(-p)=\left( -1\right) ^{J}F_{s_{1}s_{2}}(p)$, so that $%
P=\left( -1\right) ^{J+1}$.

\noindent For the triplet states with\ $\ell =J$ we get from (38) $%
F_{s_{1}s_{2}}(-p)=\left( -1\right) ^{J}F_{s_{1}s_{2}}(p)$, hence $P=\left(
-1\right) ^{J+1}$.

\noindent For the triplet states with\ $\ell =J\pm 1$ we get from (39) $%
F_{s_{1}s_{2}}(-p)=\left( -1\right) ^{J+1}F_{s_{1}s_{2}}(p)$, therefore $%
P=\left( -1\right) ^{J}$.

Charge conjugation is associated with the interchange of the particle and
antiparticle. Applying the charge conjugation operator to the trial state
(20) we get 
\begin{eqnarray}
\widehat{\mathcal{C}}\left| \psi _{T}\right\rangle &=&\underset{s_{1}s_{2}}{%
\sum }\int d^{3}\mathbf{p}F_{s_{1}s_{2}}(\mathbf{p})\widehat{\mathcal{C}}%
b_{ps_{1}}^{\dagger }d_{-ps_{2}}^{\dagger }\left| 0\right\rangle \\
&=&\underset{s_{1}s_{2}}{\sum }\int d^{3}\mathbf{p}F_{s_{1}s_{2}}(\mathbf{p})%
\widehat{\mathcal{C}}b_{ps_{1}}^{\dagger }\widehat{\mathcal{C}}^{-1}\widehat{%
\mathcal{C}}d_{-ps_{2}}^{\dagger }\widehat{\mathcal{C}}^{-1}\widehat{%
\mathcal{C}}\left| 0\right\rangle .
\end{eqnarray}
Using the relations 
\begin{equation}
\widehat{\mathcal{C}}b_{ps_{1}}^{\dagger }\widehat{\mathcal{C}}^{-1}=\eta
^{C}d_{ps_{1}}^{\dagger },\;\;\;\;\;\widehat{\mathcal{C}}d_{-ps_{2}}^{%
\dagger }\widehat{\mathcal{C}}^{-1}=\eta ^{C}b_{-ps_{2}}^{\dagger
},\;\;\;\;\;\widehat{\mathcal{C}}\left| 0\right\rangle =\left|
0\right\rangle ,
\end{equation}
where $\left( \eta ^{C}\right) ^{2}=1$, we obtain 
\begin{equation*}
\widehat{\mathcal{C}}\left| \psi _{T}\right\rangle =\underset{s_{1}s_{2}}{%
\sum }\int d^{3}\mathbf{p}F_{s_{1}s_{2}}(\mathbf{p})\widehat{\mathcal{C}}%
b_{ps_{1}}^{\dagger }d_{-ps_{2}}^{\dagger }\left| 0\right\rangle =-\underset{%
s_{1}s_{2}}{\sum }\int d^{3}\mathbf{p}F_{s_{2}s_{1}}(\mathbf{p}%
)b_{ps_{1}}^{\dagger }d_{-ps_{2}}^{\dagger }\left| 0\right\rangle
\end{equation*}
\begin{equation}
=C\underset{s_{1}s_{2}}{\sum }\int d^{3}\mathbf{p}F_{s_{1}s_{2}}(\mathbf{p}%
)b_{ps_{1}}^{\dagger }d_{-ps_{2}}^{\dagger }\left| 0\right\rangle ,
\end{equation}
where the charge conjugation quantum number $C$ depends on the symmetry of $%
F_{s_{1}s_{2}}(p)$ in different states:

\noindent For the singlet states\ $\left( \ell =J\right) $ we get from (36) $%
F_{s_{1}s_{2}}(-p)=\left( -1\right) ^{J+1}F_{s_{1}s_{2}}(p)$, hence $%
C=\left( -1\right) ^{J}$.

\noindent For the triplet states with\ $\ell =J$ we get from (38) $%
F_{s_{1}s_{2}}(-p)=\left( -1\right) ^{J}F_{s_{1}s_{2}}(p)$, therefore $%
C=\left( -1\right) ^{J+1}$.

\noindent For the triplet states with\ $\ell =J\pm 1$ we get from (39) $%
F_{s_{1}s_{2}}(-p)=\left( -1\right) ^{J+1}F_{s_{1}s_{2}}(p)$, so that $%
C=\left( -1\right) ^{J}$.


\vskip0.8truecm \noindent{\textbf{\large Appendix C. Expansion of the spinors%
}}

\vskip0.4truecm


We recall the form of the particle spinors: 
\begin{equation}
u(\mathbf{p,}i)=N_{\mathbf{p}}\left[ 
\begin{array}{c}
1 \\ 
\frac{(\overrightarrow{\sigma }\mathbf{\cdot p})}{\omega _{p}+m}
\end{array}
\right] \varphi _{i},
\end{equation}
where 
\begin{equation}
\varphi _{1}=\left[ 
\begin{array}{c}
1 \\ 
0
\end{array}
\right] ,\;\;\;\varphi _{2}=\left[ 
\begin{array}{c}
0 \\ 
1
\end{array}
\right] ,\;\;\;\;\;N_{\mathbf{p}}=\sqrt{\frac{\omega _{p}+m}{2m}}.
\end{equation}
The antiparticle or ``positron'' representation for the $v_{i}(p)$ spinors
has the form

\begin{equation}
v(\mathbf{p,}i)=N_{\mathbf{p}}\left[ 
\begin{array}{c}
\frac{(\overrightarrow{\sigma }\mathbf{\cdot p})}{\omega _{p}+m} \\ 
1
\end{array}
\right] \chi _{i},
\end{equation}
where 
\begin{equation}
\chi _{1}=\left[ 
\begin{array}{c}
0 \\ 
1
\end{array}
\right] ,\;\;\;\;\;\chi _{2}=-\left[ 
\begin{array}{c}
1 \\ 
0
\end{array}
\right] .
\end{equation}
The normalization is 
\begin{equation}
\overline{u}(\mathbf{p,}i)u(\mathbf{p,}j)=\delta _{ij},\;\;\;\;\;\;\;%
\overline{v}(\mathbf{p,}i)v(\mathbf{p,}j)=-\delta _{ij}.
\end{equation}
Expanding in powers of $p/m$ and keeping the lowest non-trivial order terms,

\begin{equation}
\frac{(\overrightarrow{\sigma }\mathbf{\cdot p})}{\omega _{p}+m}\simeq \frac{%
(\overrightarrow{\sigma }\mathbf{\cdot p})}{2m},
\end{equation}
\begin{equation}
N_{\mathbf{p}}=\sqrt{\frac{\omega _{p}+m}{2m}}\simeq 1+\frac{\mathbf{p}^{2}}{%
8m^{2}},
\end{equation}
we obtain the result 
\begin{equation}
u(\mathbf{p,}i)\simeq \left( 1+\frac{\mathbf{p}^{2}}{8m^{2}}\right) \left[ 
\begin{array}{c}
1 \\ 
\frac{(\overrightarrow{\sigma }\mathbf{\cdot p})}{2m}
\end{array}
\right] \varphi _{i}=\left[ 
\begin{array}{c}
\left( 1+\frac{\mathbf{p}^{2}}{8m^{2}}\right) \\ 
\frac{(\overrightarrow{\sigma }\mathbf{\cdot p})}{2m}
\end{array}
\right] \varphi _{i},
\end{equation}
\begin{equation}
v(\mathbf{p,}i)\simeq \left( 1+\frac{\mathbf{p}^{2}}{8m^{2}}\right) \left[ 
\begin{array}{c}
\frac{(\overrightarrow{\sigma }\mathbf{\cdot p})}{2m} \\ 
1
\end{array}
\right] \chi _{i}=\left[ 
\begin{array}{c}
\frac{(\overrightarrow{\sigma }\mathbf{\cdot p})}{2m} \\ 
\left( 1+\frac{\mathbf{p}^{2}}{8m^{2}}\right)
\end{array}
\right] \chi _{i}.
\end{equation}


\vskip0.8truecm \noindent{\textbf{\large Appendix D. Some useful identities
and integrals}}

\vskip0.4truecm


The following identity is useful for evaluating the $\mathcal{M}$ matrices: 
\begin{equation}
\frac{\left( \left( \mathbf{p}-\mathbf{q}\right) \cdot \mathbf{p}\right) ^{2}%
}{\left( \mathbf{p}-\mathbf{q}\right) ^{4}}=\frac{\mathbf{p}^{2}}{\left( 
\mathbf{p}-\mathbf{q}\right) ^{2}}-\frac{\left( \mathbf{p}\times \mathbf{q}%
\right) ^{2}}{\left( \mathbf{p}-\mathbf{q}\right) ^{4}}.
\end{equation}

The angular integration in (47), (50), (55) involves the following integrals

{\normalsize 
\begin{equation}
\int d\hat{\mathbf{p}}\,d\hat{\mathbf{q}}\,\digamma \left( \hat{\mathbf{p}}%
\cdot \hat{\mathbf{q}}\right) Y_{J^{\prime }}^{m_{J}^{\prime }}(\hat{\mathbf{%
q}})Y_{J}^{m_{J}\ast }(\hat{\mathbf{p}})=2\pi \delta _{J^{\prime }J}\delta
_{m_{J}^{\prime }m_{J}}\int d\left( \hat{\mathbf{p}}\cdot \hat{\mathbf{q}}%
\right) \digamma \left( \hat{\mathbf{p}}\cdot \hat{\mathbf{q}}\right)
P_{J}\left( \hat{\mathbf{p}}\cdot \hat{\mathbf{q}}\right) ,
\end{equation}
}

{\normalsize 
\begin{equation}
\int d\left( \hat{\mathbf{p}}\cdot \hat{\mathbf{q}}\right) \frac{\hat{%
\mathbf{p}}\cdot \hat{\mathbf{q}}}{\left( \mathbf{p}-\mathbf{q}\right) ^{2}}%
P_{J}\left( \hat{\mathbf{p}}\cdot \hat{\mathbf{q}}\right) =\frac{1}{\left| 
\mathbf{p}\right| \left| \mathbf{q}\right| }\left( \frac{J+1}{2J+1}%
Q_{J+1}\left( z\right) +\frac{J}{2J+1}Q_{J-1}\left( z\right) \right) ,
\end{equation}
}

\begin{equation}
\int d\left( \hat{\mathbf{p}}\cdot \hat{\mathbf{q}}\right) \frac{\left( 
\mathbf{p}\times \mathbf{q}\right) ^{2}}{\left( \mathbf{p}-\mathbf{q}\right)
^{4}}P_{J}\left( \hat{\mathbf{p}}\cdot \hat{\mathbf{q}}\right) =\frac{\left(
J+1\right) \left( J+2\right) }{2\left( 2J+1\right) }Q_{J+1}\left( z\right) -%
\frac{J\left( J-1\right) }{2\left( 2J+1\right) }Q_{J-1}\left( z\right) ,
\end{equation}
where $\digamma \left( \hat{\mathbf{p}}\cdot \hat{\mathbf{q}}\right) $ is an
arbitrary function of $\hat{\mathbf{p}}\cdot \hat{\mathbf{q}}$, $P_{J}\left(
x\right) $ is the Legendre polynomial, and $Q_{J}(z)$ is the Legendre
function of the second kind of order $J$.

The following integrals are needed for the calculation of the relativistic
energy corrections. 
\begin{equation}
\int_{0}^{\infty }\int_{0}^{\infty }dp\,dq\,p^{2}q^{2}f^{J}(p)f^{J}(q)=2\pi
\left( \frac{\alpha \mu }{n}\right) ^{3}\delta _{J,0},
\end{equation}
\begin{equation}
\int_{0}^{\infty }\int_{0}^{\infty }dp\,dq\,pqf^{J}(p)f^{J}(q)Q_{J}(z_{1})=%
\frac{\pi \alpha \mu }{n^{2}},
\end{equation}
\begin{eqnarray}
\int_{0}^{\infty }\int_{0}^{\infty
}dp\,dq\,p^{2}q^{2}f^{J}(p)f^{J}(q)Q_{J}(z_{1}) &=&  \notag \\
\int_{0}^{\infty }\int_{0}^{\infty
}dp\,dq\,p^{3}qf^{J}(p)f^{J}(q)Q_{J}(z_{1}) &=&\pi \left( \frac{\alpha \mu }{%
n}\right) ^{3}\left( \frac{4}{2J+1}-\frac{1}{n}\right) ,
\end{eqnarray}
\begin{equation}
\int_{0}^{\infty }\int_{0}^{\infty
}dp\,dq\,p^{2}q^{2}f^{J}(p)f^{J}(q)Q_{J-1}(z_{1})=\pi \left( \frac{\alpha
\mu }{n}\right) ^{3}\left( \frac{2}{J}-\frac{1}{n}\right) ,
\end{equation}
\begin{equation}
\int_{0}^{\infty }\int_{0}^{\infty
}dp\,dq\,p^{2}q^{2}f^{J}(p)f^{J}(q)Q_{J+1}\left( z_{1}\right) =\pi \left( 
\frac{\alpha \mu }{n}\right) ^{3}\left( \frac{2}{J+1}-\frac{1}{n}\right) .
\end{equation}
Here $f^{J}$ is the nonrelativistic hydrogen-like\ radial wave function in
momentum space [19]

\begin{equation}
f^{J}(p)\equiv f_{n}^{J}(p)=\left( \frac{2}{\pi }\frac{\left( n-J-1\right) !%
}{\left( n+J\right) !}\right) ^{1/2}\frac{n^{J+2}p^{J}2^{2\left( J+1\right)
}J!}{\left( n^{2}p^{2}+1\right) ^{J+2}}\mathcal{G}_{n-J-1}^{J+1}\left( \frac{%
n^{2}p^{2}-1}{n^{2}p^{2}+1}\right) ,
\end{equation}
where $G_{n-J-1}^{J+1}\left( x\right) $ are Gegenbauer functions.


\vskip0.8truecm \noindent{\textbf{{\large Appendix E. $\mathcal{K}_{12}$, $%
\mathcal{K}_{21}$ kernels for \ $l=J\mp 1$\ states}}}

\vskip0.4truecm


The contribution of the kernel $K_{12}$ to the energy correction is 
\begin{equation}
\int dp\,dq\,p^{2}q^{2}\mathcal{K}_{12}\left( p,q\right)
f^{J-1}(p)f^{J+1}(q),
\end{equation}
where

\begin{equation}
\mathcal{K}_{12}\left( p,q\right) =\underset{\sigma _{1}\sigma _{2}s_{1}s_{2}%
}{\sum }C_{Jm_{J}12}^{s_{1}s_{2}\sigma _{1}\sigma _{2}}\int d\hat{\mathbf{p}}%
\,d\hat{\mathbf{q}}\,\mathcal{M}_{s_{1}s_{2}\sigma _{1}\sigma
_{2}}^{ope\left( s-s\right) }\left( \mathbf{p,q}\right) Y_{J+1}^{m_{\sigma
_{1}\sigma _{2}}}(\hat{\mathbf{q}})Y_{J-1}^{m_{s_{1}s_{2}}\ast }(\hat{%
\mathbf{p}}).
\end{equation}
This requires the following integral 
\begin{equation}
\underset{\sigma _{1}\sigma _{2}s_{1}s_{2}}{\sum }C_{Jm_{J}12}^{s_{1}s_{2}%
\sigma _{1}\sigma _{2}}\int d^{3}\mathbf{p\,}d^{3}\mathbf{q\,}%
f^{J-1}(p)Y_{J-1}^{m_{s_{1}s_{2}}\ast }(\hat{\mathbf{p}})\mathcal{M}%
_{s_{1}s_{2}\sigma _{1}\sigma _{2}}^{ope\left( s-s\right) }\left( \mathbf{p,q%
}\right) f^{J+1}(q)Y_{J+1}^{m_{\sigma _{1}\sigma _{2}}}(\hat{\mathbf{q}}).
\end{equation}
We calculate this form in coordinate space. The Fourier transform of $%
M_{s_{1}s_{2}\sigma _{1}\sigma _{2}}\left( \mathbf{p,q}\right) \;$is

\begin{equation}
\mathcal{M}_{s_{1}s_{2}\sigma _{1}\sigma _{2}}\left( \mathbf{p,q}\right)
=\int d^{3}\mathbf{r}\,d^{3}\mathbf{r}^{\prime }\mathcal{M}%
_{s_{1}s_{2}\sigma _{1}\sigma _{2}}\left( \mathbf{r,r}^{\prime }\right)
e^{-i\left( \mathbf{p-q}\right) \cdot \left( \mathbf{r-r}^{\prime }\right) },
\end{equation}
where the $M_{s_{1}s_{2}\sigma _{1}\sigma _{2}}\left( \mathbf{r,r}^{\prime
}\right) $ matrix is a local operator in general [16], that is 
\begin{equation}
\mathcal{M}_{s_{1}s_{2}\sigma _{1}\sigma _{2}}\left( \mathbf{r,r}^{\prime
}\right) =\mathcal{M}_{s_{1}s_{2}\sigma _{1}\sigma _{2}}\left( \mathbf{r}%
\right) \delta \left( \mathbf{r-r}^{\prime }\right) .
\end{equation}
We apply this transformation to the $M_{s_{1}s_{2}\sigma _{1}\sigma
_{2}}^{ope\left( s-s\right) }\left( \mathbf{p,q}\right) $ matrix (see eq.
(62)). Because of the angular integration in (149), only the first term in
(62) survives. The\ Fourier transformation of that term is 
\begin{equation}
\frac{\left( \overrightarrow{\sigma }^{\left( +\right) }\cdot \left( \mathbf{%
p-q}\right) \right) \left( \overrightarrow{\sigma }^{\left( -\right) }\cdot
\left( \mathbf{p-q}\right) \right) }{4m^{2}\left( \mathbf{p-q}\right) ^{2}}%
\;\;\;\rightarrow \;\;\;3\frac{\left( \overrightarrow{\sigma }^{\left(
+\right) }\cdot \mathbf{r}\right) \left( \overrightarrow{\sigma }^{\left(
-\right) }\cdot \mathbf{r}\right) }{16\pi m^{2}r^{5}}.
\end{equation}
Furthermore,

\begin{equation}
\int d^{3}\mathbf{p}f^{J-1}(p)Y_{J-1}^{m_{s_{1}s_{2}}\ast }(\hat{\mathbf{p}}%
)e^{-i\mathbf{p}\cdot \mathbf{r}}=R_{n}^{J-1}(r)Y_{J-1}^{m_{s_{1}s_{2}}\ast
}(\hat{\mathbf{r}}),
\end{equation}
\begin{equation}
\int d^{3}\mathbf{q}f^{J+1}(q)Y_{J+1}^{m_{s_{1}s_{2}}\ast }(\hat{\mathbf{q}}%
)e^{-i\mathbf{q}\cdot \mathbf{r}}=R_{n}^{J+1}(r)Y_{J+1}^{m_{s_{1}s_{2}}}(%
\hat{\mathbf{r}}),
\end{equation}
where 
\begin{equation}
R_{n}^{\ell }\left( r\right) =-\frac{2}{n^{2}}\sqrt{\frac{\left( n-\ell
-1\right) !}{\left( \left( n+\ell \right) !\right) ^{3}}}e^{-r/n}\left( 
\frac{2r}{n}\right) ^{\ell }L_{n+\ell }^{2\ell +1}\left( \frac{2r}{n}\right)
.
\end{equation}
The associated Laguerre function \ \ $L_{\lambda }^{\mu }\left( \rho \right) 
$\ \ is related to the confluent hypergeometric function by 
\begin{equation}
L_{\lambda }^{\mu }\left( \rho \right) =\left( -1\right) ^{\mu }\frac{\left(
\lambda !\right) ^{2}}{\mu !\left( \lambda -\mu \right) !}F\left( -\lambda
+\mu ,\mu +1;\rho \right) .
\end{equation}
The generating function for the Laguerre function is 
\begin{equation}
U_{\mu }\left( \rho ,u\right) \equiv \left( -1\right) ^{\mu }\frac{u^{\mu }}{%
\left( 1-u\right) ^{\mu +1}}\exp \left( -\frac{u\rho }{1-u}\right)
=\sum_{\lambda =\mu }^{\infty }\frac{L_{\lambda }^{\mu }\left( \rho \right) 
}{\lambda !}u^{\lambda },
\end{equation}
hence 
\begin{eqnarray}
&&\underset{\sigma _{1}\sigma _{2}s_{1}s_{2}}{\sum }C_{Jm_{J}12}^{s_{1}s_{2}%
\sigma _{1}\sigma _{2}}\int d^{3}\mathbf{p\,}d^{3}\mathbf{q\,}%
f^{J-1}(p)Y_{J-1}^{m_{s_{1}s_{2}}\ast }(\hat{\mathbf{p}})\mathcal{M}%
_{s_{1}s_{2}\sigma _{1}\sigma _{2}}^{ope\left( s-s\right) }\left( \mathbf{p,q%
}\right) f^{J+1}(q)Y_{J+1}^{m_{\sigma _{1}\sigma _{2}}}(\hat{\mathbf{q}}) 
\notag \\
&=&\underset{\sigma _{1}\sigma _{2}s_{1}s_{2}}{\sum }%
C_{Jm_{J}12}^{s_{1}s_{2}\sigma _{1}\sigma _{2}}\int d^{3}\mathbf{r\,}%
R_{n}^{J-1}\left( r\right) Y_{J-1}^{m_{s_{1}s_{2}}\ast }(\hat{\mathbf{r}}) 
\notag \\
&&\times \left( 3\alpha \frac{\left( \overrightarrow{\sigma }^{\left(
+\right) }\cdot \mathbf{r}\right) \left( \overrightarrow{\sigma }^{\left(
-\right) }\cdot \mathbf{r}\right) }{16\pi m^{2}r^{5}}\right)
R_{n}^{J+1}\left( r\right) Y_{J+1}^{m_{s_{1}s_{2}}}(\hat{\mathbf{r}}) \\
&=&\frac{3\alpha }{16\pi m^{2}}\int dr\,r^{2}\frac{1}{r^{3}}%
R_{n}^{J-1}\left( r\right) R_{n}^{J+1}\left( r\right) \times  \notag \\
&&\times \underset{\sigma _{1}\sigma _{2}s_{1}s_{2}}{\sum }%
C_{Jm_{J}12}^{s_{1}s_{2}\sigma _{1}\sigma _{2}}\int d\hat{\mathbf{r}}%
\,Y_{J-1}^{m_{s_{1}s_{2}}\ast }(\hat{\mathbf{r}})\left( \overrightarrow{%
\sigma }^{\left( +\right) }\cdot \hat{\mathbf{r}}\right) \left( 
\overrightarrow{\sigma }^{\left( -\right) }\cdot \hat{\mathbf{r}}\right)
Y_{J+1}^{m_{s_{1}s_{2}}}(\hat{\mathbf{r}}).  \notag
\end{eqnarray}
It follows that 
\begin{equation}
\underset{\sigma _{1}\sigma _{2}s_{1}s_{2}}{\sum }C_{Jm_{J}12}^{s_{1}s_{2}%
\sigma _{1}\sigma _{2}}\int d\hat{\mathbf{r}}Y_{J-1}^{m_{s_{1}s_{2}}\ast }(%
\hat{\mathbf{r}})\left( \overrightarrow{\sigma }^{\left( +\right) }\cdot 
\hat{\mathbf{r}}\right) \left( \overrightarrow{\sigma }^{\left( -\right)
}\cdot \hat{\mathbf{r}}\right) Y_{J+1}^{m_{s_{1}s_{2}}}(\hat{\mathbf{r}})=%
\frac{1}{15}\frac{\sqrt{J\left( J+1\right) }}{2J+1},
\end{equation}
but 
\begin{equation}
\int_{0}^{\infty }dr\,r^{2}\frac{1}{r^{3}}R_{n}^{J-1}\left( r\right)
R_{n}^{J+1}\left( r\right) =0.
\end{equation}
The last expression can be proved in the following way. Let us consider the
more general case 
\begin{equation}
\int_{0}^{\infty }dr\,r^{\beta +2}R_{n}^{\ell }\left( r\right) R_{n}^{\ell
^{\prime }}\left( r\right) .
\end{equation}
The generating function for $R_{n}^{\ell }\left( r\right) $\ is 
\begin{equation}
G_{n\ell }\left( r,u\right) =-\frac{2}{n^{2}}\sqrt{\frac{\left( n-\ell
-1\right) !}{\left( \left( n+\ell \right) !\right) ^{3}}}e^{-r/n}\left( 
\frac{2r}{n}\right) ^{\ell }\left( -1\right) ^{2\ell +1}\frac{u^{2\ell +1}}{%
\left( 1-u\right) ^{2\ell +2}}\exp \left\{ -\frac{u}{1-u}\frac{2r}{n}%
\right\} .
\end{equation}
Then we consider the expression 
\begin{eqnarray}
&&\int_{0}^{\infty }drr^{\beta +2}G_{n\ell }\left( r,u\right) G_{n\ell
^{\prime }}\left( r,v\right)  \notag \\
&=&\int_{0}^{\infty }drr^{\beta +2}\frac{4}{n^{4}}\sqrt{\frac{\left( n-\ell
-1\right) !\left( n-\ell ^{\prime }-1\right) !}{\left( \left( n+\ell \right)
!\right) ^{3}\left( \left( n+\ell ^{\prime }\right) !\right) ^{3}}}%
e^{-2r/n}\left( \frac{2r}{n}\right) ^{\ell +\ell ^{\prime }}\times  \notag \\
&&\times \frac{u^{2\ell +1}v^{2\ell ^{\prime }+1}}{\left( 1-u\right) ^{2\ell
+2}\left( 1-v\right) ^{2\ell ^{\prime }+2}}\exp \left\{ -\left( \frac{u}{1-u}%
+\frac{v}{1-v}\right) \frac{2r}{n}\right\} \\
&=&\frac{4}{n^{4}}\sqrt{\frac{\left( n-\ell -1\right) !\left( n-\ell
^{\prime }-1\right) !}{\left( \left( n+\ell \right) !\right) ^{3}\left(
\left( n+\ell ^{\prime }\right) !\right) ^{3}}}\frac{u^{2\ell +1}v^{2\ell
^{\prime }+1}}{\left( 1-u\right) ^{2\ell +2}\left( 1-v\right) ^{2\ell
^{\prime }+2}}\times  \notag \\
&&\times \int_{0}^{\infty }dr\left( \frac{2r}{n}\right) ^{\beta +2+\ell
+\ell ^{\prime }}\exp \left\{ -\left( 1+\frac{u}{1-u}+\frac{v}{1-v}\right) 
\frac{2r}{n}\right\} .  \notag
\end{eqnarray}
It is well known that 
\begin{equation}
\int_{0}^{\infty }d\rho \rho ^{\beta }e^{-\rho }=\Gamma \left( \beta
+1\right) ,
\end{equation}
therefore 
\begin{eqnarray}
&&\int_{0}^{\infty }dr\left( \frac{2r}{n}\right) ^{\beta +2+\ell +\ell
^{\prime }}\exp \left\{ -\left( 1+\frac{u}{1-u}+\frac{v}{1-v}\right) \frac{2r%
}{n}\right\}  \notag \\
&=&\left( \frac{n}{2}\right) ^{\beta +3}\left( \frac{\left( 1-u\right)
\left( 1-v\right) }{1-uv}\right) ^{\beta +3+\ell +\ell ^{\prime }}\Gamma
\left( \beta +3+\ell +\ell ^{\prime }\right)
\end{eqnarray}
and 
\begin{eqnarray}
&&\int_{0}^{\infty }drr^{\beta +2}G_{n\ell }\left( r,u\right) G_{n\ell
^{\prime }}\left( r,v\right)  \notag \\
&=&\frac{2^{-\beta -1}}{n^{-\beta +1}}\sqrt{\frac{\left( n-\ell -1\right)
!\left( n-\ell ^{\prime }-1\right) !}{\left( \left( n+\ell \right) !\right)
^{3}\left( \left( n+\ell ^{\prime }\right) !\right) ^{3}}}\times \\
&&\times \frac{u^{2\ell +1}v^{2\ell ^{\prime }+1}\left( 1-u\right) ^{\beta
+1-\ell +\ell ^{\prime }}\left( 1-v\right) ^{\beta +1+\ell -\ell ^{\prime }}%
}{\left( 1-uv\right) ^{\beta +3+\ell +\ell ^{\prime }}}\Gamma \left( \beta
+3+\ell +\ell ^{\prime }\right) .  \notag
\end{eqnarray}
We expand this expression in a series, 
\begin{equation}
\int_{0}^{\infty }drr^{\beta +2}G_{n\ell }\left( r,u\right) G_{n\ell
^{\prime }}\left( r,v\right) =\sum_{\eta \eta ^{\prime }}C_{\eta \eta
^{\prime }}\left( n,\beta ,\ell ,\ell ^{\prime }\right) u^{\eta }u^{\eta
^{\prime }}.
\end{equation}
It is not difficult to show [20], that the coefficient $C_{n+\ell ,n+\ell
^{\prime }}$\ represents the integral\ 
\begin{equation}
C_{n+\ell ,n+\ell ^{\prime }}\left( n,\beta ,\ell ,\ell ^{\prime }\right)
=\int_{0}^{\infty }drr^{\beta +2}R_{n}^{\ell }(r)R_{n}^{\ell ^{\prime }}(r).
\end{equation}
Simple but tedious calculations show that this\ coefficient is zero for $%
\beta =-3$, $\ell =J-1$, $\ell ^{\prime }=J+1$. Thus the kernel $K_{12}$
does not contribute to the energy corrections to $O\left( \alpha ^{4}\right) 
$. The same result is obtained for the kernel $K_{21}$.

{\normalsize 
}

{\normalsize \vskip0.8truecm \noindent{\textbf{\large References}} }

{\normalsize \vskip0.4truecm }

{\normalsize 
}

{\normalsize \enumerate}

1. J. W. Darewych, Annales Fond. L. de Broglie (Paris) \textbf{23}, 15
(1998).

2. J. W. Darewych, in \textit{Causality and Locality in Modern Physics}, G
Hunter et al. (eds.), Kluwer, Dordrecht, p. 333, (1998).

3. J. W. Darewych, Can. J. Phys. \textbf{76}, 523 (1998).

4. M. Barham and J. W. Darewych, J. Phys. A \textbf{31}, 3481 (1998).

5. B. Ding and J. Darewych, J. Phys. G \textbf{26}, 907 (2000).

6. J. D. Jackson, \textit{Classical Electrodynamics} (John Wiley, New York,
1975).

7. A. O. Barut, \textit{Electrodynamics and Classical Theory of Fields and
Particles} (Dover, new York, 1980).

8. W. T. Grandy, \textit{Relativistic Quantum Mechanics of Leptons and Fields%
} (Kluwer, Dordrecht, 1991).

9. A. O. Barut, in \textit{Geometrical and Algebraic Aspects of Nonlinear
Field Theory}, edited by S. De Filippo, M. Marinaro, G. Marmo and G. Vilasi,
(Elsevier New York, 1989), p. 37.

10. J. W. Darewych and L. Di Leo, J. Phys. A \textbf{29}, 6817 (1996).

11. H. Pilkuhn, J. Phys. B \textbf{28,} 4421 (1995), H. Pilkuhn, \textit{%
Relativistic Quantum Mechanics} (Springer 2003).

12. J. W. Darewych and M. Horbatsch, J. Phys. B \textbf{22}, 973 (1989); 
\textbf{23}, 337 (1990).

13. W. Dykshoorn and R. Koniuk, Phys. Rev. A \textbf{41}, 64 (1990).

14. T. Zhang and R. Koniuk, Can. J. Phys. \textbf{70}, 683 (1992).

15. T. Zhang and R. Koniuk, Can. J. Phys. \textbf{70}, 670 (1992).

16. V. B. Berestetskii, E. M. Lifshitz, and L. P. Pitaevski, \textit{Quantum
Electrodynamics} (Pergamon, New York, 1982).

17. G. Arfken and H. Weber, \textit{Mathematical Methods for Physicists}
(Academic Press, 2001).

18. M. A. Stroscio, Positronium: A Review Of The Theory, Physics Reports C 
\textbf{22}, 5-215 (1975).

19. H. A. Bethe and E. E. Salpeter, \textit{Quantum Mechanics of One and
Two-Electron Atoms} (Springer-Verlag/Academic, New York, 1957).

20. M. Mizushima, \textit{Quantum Mechanics of Atomic Spectra and Atomic
Structure} (W. A. Benjamin, Inc., New York, 1970).

\end{document}